\documentclass[12pt,preprint]{aastex}

\newcommand{\G}{\Gamma}

\newcommand{\sT}{\sigma_{\rm T}}
\newcommand{\pr}{^\prime}
\newcommand{\e}{\epsilon}
\newcommand{\g}{\gamma}
\newcommand{\gp}{\gamma^{\prime}}

\newcommand{\mup}{\mu^\prime}

\newcommand{\ep}{\epsilon^\prime}

\newcommand{\dD}{\delta_{\rm D}}

\newcommand{\psim}{\lower.5ex\hbox{$\; \buildrel \propto \over\sim \;$}}
\newcommand{\lbar}{\lower.0ex\hbox{$\; \buildrel 
{\lower0.0ex \hbox{-}} \over\lambda  \;$}}

\shorttitle{$\gamma$-ray XBL Blazar Analysis}
\shortauthors{Finke, Dermer, and B\"ottcher}

\begin{document}
\title {Synchrotron Self-Compton Analysis 
of TeV X-ray Selected BL Lacertae Objects   }

\author{Justin D. Finke\altaffilmark{1,2}, Charles D.\ Dermer\altaffilmark{1},  \& 
	Markus B\"ottcher\altaffilmark{3}}

\email{justin.finke@nrl.navy.mil}

\altaffiltext{1}{U.S.\ Naval Research Laboratory, Code 7653, 4555 Overlook SW, 
	Washington, DC
  	20375-5352}
\altaffiltext{2}{NRL/NRC Research Associate}
\altaffiltext{3}{Astrophysical Institute, Department of Physics and Astronomy, 
     Ohio University, Athens, Ohio 45701}

\begin{abstract}


We introduce a methodology for analysis of multiwavelength data from
X-ray selected BL Lac (XBL) objects detected in the TeV regime.  By
assuming that the radio--through--X-ray flux from XBLs is nonthermal
synchrotron radiation emitted by isotropically-distributed electrons
in the randomly oriented magnetic field of a relativistic blazar jet,
we obtain the electron spectrum.  This spectrum is then used to deduce
the synchrotron self-Compton (SSC) spectrum as a function of the
Doppler factor, magnetic field, and variability timescale.  The
variability timescale is used to infer the comoving blob radius from
light travel-time arguments, leaving only two parameters.  With this
approach, we accurately simulate the synchrotron and SSC spectra of
flaring XBLs in the Thomson through Klein-Nishina regimes.
Photoabsorption by interactions with internal jet radiation and the
intergalactic background light (IBL) is included.  Doppler factors,
magnetic fields, and absolute jet powers are obtained by fitting the
{\em HESS} and {\em Swift} data of the recent giant TeV flare observed
from \object{PKS 2155--304}.  For the {\em HESS} and {\em Swift}
data from 28 and 30 July 2006, respectively, Doppler
factors $\gtrsim 60$ and absolute jet powers $\gtrsim 10^{46}$ ergs
s$^{-1}$ are required for a synchrotron/SSC model to give a good fit
to the data, for a low intensity of the IBL and a ratio of 10 times
more energy in hadrons than nonthermal electrons.  Fits are also made
to a TeV flare observed in  2001 from \object{Mkn 421} which
require Doppler factors $\ga 30$ and jet powers $\ga 10^{45}$ erg
s$^{-1}$. 
\end{abstract}


\section{INTRODUCTION}

Radiation from blazars is thought to originate from a relativistic
jet, closely aligned with our line of sight, that is powered by a
supermassive black hole at the nucleus of an active galaxy.  Blazars
exhibit strong, rapidly-varying emission throughout the
electromagnetic spectrum from radio to $\gamma$-rays.  The broadband
spectral energy distributions (SEDs) of blazars consist of two
components \citep[e.g.\ ][]{mukherjee97,weekes03}: a low energy
component that peaks at infrared or optical energies, and a high
energy component that peaks at MeV--GeV $\gamma$-ray energies, and
often extends to very high-energy (VHE) $\g$-rays.  BL Lacertae (BL
Lac) objects are a subclass of blazars distinguished by their weakness
or absence of broad emission lines in a quasar-like optical spectrum.
BL Lac objects are often further sub-divided into those that have low
energy components peaking in the optical (known as low frequency
peaked or radio selected BL Lacs [RBLs]) and those that have low
energy components peaking in the X-rays (high frequency peaked or
X-ray selected BL Lacs [XBLs]).

There are two broad classes of models which are used to describe these
observations: leptonic models, in which the emission is primarily from
relativistic electrons and positrons; and hadronic models, in which
the emission is primarily from protons and atomic nuclei.  In the
leptonic models \citep[see, e.g., ][for a review]{boett07}, the
low-energy component is interpreted as synchrotron emission from
leptons, and the high-energy component as radiation from target
photons Compton up-scattered by relativistic jet leptons.  The target
photons can originate from the synchrotron radiation
\citep[synchrotron self-Compton, or SSC;][] {bloom96}, as well as
external sources, e.g., the broad-line region \citep{sikora94}, a
dusty torus \citep{blazejowski00}, and/or the accretion disk
\citep{dermer92,dermer93}.  In hadronic models \citep[see, e.g.,
][]{mannheim92,muecke01,atoyan03}, the low-energy component is still
interpreted as synchrotron emission from relativistic leptons, but the
high energy component originates from proton synchrotron or photopion
production initiated by interactions between protons and soft photons.
The high-energy synchrotron and Compton emissions create, following
cascading and absorption of $\gamma$-rays by the ambient photons, a
second hadronicly induced $\gamma$-ray emission component.

Blazar modeling has generally concentrated on leptonic processes to
model simultaneous multi-wavelength data \citep[e.g.,
][]{ghisellini96, li00, boett02a, boett02b, joshi07}.  In the standard
time-dependent leptonic blazar models, the low- and high-energy
components in the SEDs are simultaneously fit by injecting nonthermal
electrons (including positrons) into the jet and allowing the
electrons to evolve through radiative and adiabatic cooling.

In this paper, we use a different approach to model the SEDs of
XBLs. In \S\ 2, we fit the optical/X-ray $\nu F_\nu$ spectrum, and use
this spectral form to deduce the electron distribution in the jet
assuming that this emission is nonthermal lepton synchrotron
radiation.  We then use this electron distribution to calculate the
high-energy SSC component.  The SSC flux is then precisely given by
the inferred electron distribution and a small set of well-constrained
observables.  The full Compton cross section, accurate for
relativistic electrons from the Thomson through the Klein-Nishina
regime, is used in the derivation.  We take into account $\g\g$
absorption by high energy interactions with low energy jet radiation
and by interactions with the intergalactic background light (IBL).  We
then apply this formulation to the recently observed giant TeV flare
in the XBL \object{PKS 2155--304} \citep[\S\
3,][]{aharonian07_2155,foschini07} as well as to the 
March 2001 TeV flare observed
from \object{Mkn 421} with HEGRA and {\em RXTE} 
\citep[\S 4,][]{aharonian02,fossati08}.
Allowed values
of mean magnetic field $B$ and Doppler factor $\dD$ are obtained by
fitting the X/$\gamma$-ray spectrum of \object{PKS 2155--304} in \S\
\ref{pks2155} and \object{Mkn 421} in \S\ 4, and used to derive the apparent
isotropic jet luminosity.  The results are discussed and summarized in
\S\ 5.

\section{ANALYSIS}

Blazars vary on timescales from days to months at radio frequencies
\citep[e.g.,][]{boett03,vetal04}, and on timescales as short as a few
hours or less at X/$\gamma$-ray energies
\citep[e.g.,][]{tagliaferri03, foschini06b}.  Photoabsorption
arguments \citep{maraschi92} and observations of superluminal motions
\citep{vermeulen94} show that the emission region is moving with
relativistic speeds.

We consider a one-zone spherical blob of relativistic plasma moving with
Lorentz factor $\G = (1 - \beta^2)^{-1/2}$.  Quantities in the
observer's frame are unprimed, and quantities in the frame comoving
with the jet blob are primed, so that the comoving volume of the blob
is $V^\prime_b =4\pi R_b^{\prime 3}/3$, where $R_b^\prime$ is the
comoving radius of the emitting blob.  The angle that the jet makes
with the observer's line of sight is denoted by $\theta \equiv
\arccos\ \mu$, and the Doppler factor is given by
$\dD=[\G(1-\beta\mu)]^{-1}$.  Distinct, rapid high-energy flares observed in
blazars imply that the emitting region is confined to a small volume
with a comoving variability timescale $t_v^\prime$, limited by light
travel time, given by
\begin{equation}
\label{Rblob}
t_v^\prime \ga \frac{R_b^\prime}{c}\; .
\end{equation}
For the  observer, the measured variability timescale is 
\begin{equation}
\label{vartime}
t_v \ga t_{v,min} = \frac{(1+z)R_b^\prime}{\dD c}.
\end{equation}

In the remainder of this section, we compute the SSC spectrum using
the electron spectrum derived with the $\delta$-approximation for
synchrotron radiation (\S\ \ref{SSCdelta}).  We then derive the SSC
spectrum using a model-dependent approach to obtain the electron
spectrum by integrating over the exact synchrotron emissivity and
minimizing $\chi^2$ (\S\ \ref{SSCexact}).  We compare the
$\delta$-approximation and the full expression for synchrotron in \S\
\ref{comparison} and the Thomson and full Compton expressions in \S\
\ref{thomsoncompare}.  We determine the $\gamma\gamma$ absorption
opacity from jet radiation (\S\ \ref{photoabs}), and present
constraints based on power available in the jet (\S\ \ref{jetL}).  The
fitting technique using the exact synchrotron expression and the full
Compton cross-section is described in \S\ \ref{fittech}.  Two versions
of the Heaviside function are used: $H(x)=0$ for $x<0$ and $H(x)=1$
for $x\ge 0$; as well as $H(x; x_1, x_2)=1$ for $x_1\le x\le x_2$ and
$H(x; x_1, x_2)=0$ everywhere else.

\subsection{SSC Emission in the $\delta$-approximation for synchrotron}
\label{SSCdelta}

The $\delta$-approximation for the synchrotron spectrum is useful in
that it allows one directly to obtain the electron spectrum and then
calculate the predicted $\nu F_\nu$ SSC spectrum, $f_\e^{SSC}$, in
terms of the observed $\nu F_\nu$ synchrotron spectrum, $f_\e^{syn}$,
where $\e=h\nu/m_ec^2$ is the emitted photon's dimensionless energy
in the observer's frame. We demonstrate this approach in this
section.  The $\delta$-approximation is also used to calculate jet
power.  For detailed spectral modeling, however, a more accurate
expression is needed (\S\ \ref{SSCexact}).

The $\delta$-approximation for the synchrotron flux is
\begin{equation}
\label{fes}
f_\e^{syn} \cong {\dD^4\over 6\pi d_L^2} \; c\sigma_{\rm T} U_B 
\gamma^{\prime 3}_s N_e^\prime (\gp_s)
\end{equation}
\citep[e.g.\ ][]{dermer02}.  Here $d_L$ is the luminosity distance, $c$
is the speed of light, $\sT$ is the Thomson cross section,
$N_e^\prime(\g^\prime_s)$ is the comoving electron distribution,
\begin{equation}
\label{gamma_ps}
\gp_s =  \sqrt{{\e(1+z)\over \dD\epsilon_B}} = 
\sqrt{\e^\prime\over\epsilon_B}\;,
\end{equation}
is a synchrotron-emitting electron's Lorentz factor, and
$$
U_B = \frac{B^2}{8\pi}
$$ 
is the mean comoving magnetic-field energy density of the
randomly-oriented comoving field with mean intensity $B$.  In eq.\
(\ref{gamma_ps}), $z$ is the redshift of the object and
$\epsilon_B=B/B_{cr}$, where $B_{cr}=4.414 \times 10^{13}$ G is the
critical magnetic field.  Note that the magnetic field, $B$, is 
defined in the comoving frame, despite being unprimed.  The
$\delta$-approximation is accurate for spectral indices approximately
$2.0 \le p \le 3.5$ (where the electron distribution is
$N^\prime_e(\g^\prime ) \propto \g^{\prime -p}$).

The comoving electron distribution, $N_e^\prime(\gp_s)$, can be found
in terms of $f_\e^{syn} $ from eq.\ (\ref{fes}):
\begin{equation}
\label{Neprimegps}
N_e^\prime(\gp_s) = V^\prime_b n^\prime_e(\gp_s )\cong 
{6\pi d_L^2 f_\e^{syn}\over 
c\sigma_{\rm T} U_B\dD^4 \g^{\prime 3}_s}\;.
\end{equation}
 The synchrotron emissivity, 
$\dot n_{syn}(\e )$ (photons cm$^{-3}$ s$^{-1}$
$\e^{-1}$), is given by
\begin{equation}
\dot n^{\prime }_{syn}(\ep) \cong {2\over 3}\; { c \sigma_{\rm T} u_B }
\; \e^{\prime -1/2} \epsilon_B^{-3/2}\; n^\prime_e\left(\gp_s\right)\;,
\label{dotnsyne}
\end{equation}
with $u_B=U_B/m_ec^2$.  
Using this and eq.\ (\ref{Neprimegps}), 
one can determine the synchrotron photon number density, 
\begin{equation}
n^\prime_{syn}(\ep ) \cong {R_b^\prime\over c} \; 
\dot n_{syn}^\prime( \ep )\;\cong
 {3d_L^2 f_\e^{syn}\over 
m_ec^3R^{\prime 2}_b \dD^4 \e_B^2 \g^{\prime 4}_s}\;.
\label{nprimesynep}
\end{equation}
This can be converted to a radiation energy density through the relation
\begin{equation}
\label{uprime}
u^\prime (\ep ) = \epsilon^{\prime}m_ec^2 n^\prime_{syn}(\ep ) =  
\frac{ 3 d_L^2 f_\e^{syn} }{ cR_B^{\prime 2}\dD^4\ep }
 = \frac{ 3 d_L^2 (1+z)^2 f_\e^{syn} }{ c^3 t^2_{v,min} \dD^6 \ep }\ ,
\end{equation}
where we have made use of eqs.\ (\ref{gamma_ps}) and (\ref{vartime}).  

The SSC emissivity, integrated over volume, for isotropic and homogeneous 
photon and electron distributions is given by 
\begin{equation}
\ep_s J^\prime_{SSC}(\ep_s ) = {3\over 4} c \sigma_{\rm T} \e^{\prime 2}_s
\int_0^\infty d\ep\;{u^\prime (\ep )\over \e^{\prime 2}}
 \int_{\gp_{min}}^{\gp_{max}}d\gp\;{N_e^\prime(\gp )\over 
\g^{\prime 2}}\;F_{\rm C} (q,\Gamma_e)\;,
\label{epsjssc}
\end{equation}
where for a homogeneous distribution, 
$\ep_s J^\prime_{SSC}(\ep_s ) = V^\prime_b \ep_s j^\prime_{SSC}(\ep_s )$ and 
$\ep_s $ is the scattered photon's dimensionless energy in the blob frame.
In eq.\ (\ref{epsjssc}) the Compton scattering kernel 
for isotropic photon and electron distributions is 
\begin{eqnarray}
F_{\rm C}(q, \Gamma_e)  =  \biggr[ 2q \ln q +(1+2q)(1-q) + 
{1\over 2} {(\Gamma_e q)^2\over (1+\Gamma_e q)}(1-q) \biggr]
\;H\;\left( q; {1\over 4\gamma^{\prime 2}}, 1 \right)
\label{jesC}
\end{eqnarray}
\citep{jones68,blumen70}, where
\begin{equation}
q \equiv {\ep_s/\gp \over \Gamma_e
(1-\ep_s/\gp )}\;\;{\rm and}\; \; \Gamma_e = 4\ep\gp\; .
\label{jesCq}
\end{equation}
The limits on $q$ are
\begin{equation}
\frac{1}{4\g^{\prime 2}}\ \le\ q\ \le 1\ ,
\end{equation}
which imply the limits of the integration over $\gp$:
\begin{equation}
\gp_{min} = {1\over 2} \ep_s\;\left( 1+\sqrt{1+{1\over \ep\ep_s}} \right)
\label{gpmin}
\end{equation}
and
\begin{equation}
\gp_{max} = {\ep\ep_s\over \ep - \ep_s}H(\ep -  \ep_s) \;+\; 
\gp_2H(\ep_s - \ep )\ .
\label{gpmax}
\end{equation}
The upper limit, eq.\ (\ref{gpmax}), takes into account Compton up-- and 
down--scattering. In principle, the maximum accelerated electron 
energy $\gp_2$ can be determined from particle acceleration theory.

The $\nu F_\nu$ SSC spectrum is given by
\begin{equation}
f_{\e_s}^{SSC} = \frac{ \dD^4 \ep_s J^\prime_{SSC}(\ep_s) }{ 4\pi d_L^2}\ .
\label{feSSC}
\end{equation}
Inserting eq.\ (\ref{epsjssc}) into eq.\ (\ref{feSSC}) and 
using eqs. (\ref{uprime}) and (\ref{Neprimegps}) gives
\begin{eqnarray}
f_{\e_s}^{SSC} = \left( {3\over 2}\right)^3\; {d_L^2 \e_s^{\prime 2}
\over 
R_b^{\prime 2} c \dD^4 U_B}\;
\int_0^\infty d\ep\;{f_{\tilde \e}^{syn}\over \e^{\prime 3}}\;
\int_{\gp_{min}}^{\gp_{max}}\;d\gp\ {F_{\rm C}(q,\Gamma_e)f_{\hat \e}^{syn}\over 
\g^{\prime 5} }\;,
\label{feSSC_big}
\end{eqnarray}
where
$$
\tilde \e = {\dD \ep\over 1+z} \;,
$$
$$ 
\hat \e = {\dD \e_B \g^{\prime 2}\over 1+z} \;,\;
$$
and
$$\ep_s = {(1+z)\e_s\over\dD}\;.$$
Using eq.\ (\ref{vartime}) to give an estimate of the blob's radius, 
$$
R^\prime_b \cong \frac{ c\dD t_{v,min} }{1+z}\ ,
$$
we obtain 
\begin{eqnarray}
f_{\e_s}^{SSC} = {27\pi\e_s^{\prime 2} \over c^3} \ 
\left( \frac{d_L}{1+z} \right)^2\ 
\left( \frac{1}{t_{v,min}\dD^3 B} \right)^2\ 
\int_0^\infty d\ep\;{f_{\tilde \e}^{syn}\over \e^{\prime 3}}\;
\int_{\gp_{min}}^{\gp_{max}}\;d\gp\; 
{F_{\rm C}(q,\Gamma_e)f_{\hat \e}^{syn}\over 
\g^{\prime 5} }\;.
\label{feSSC2}
\end{eqnarray}
The observed SSC spectrum in eq.\ (\ref{feSSC2}) is a function of
three observables, namely redshift $z$, the observed synchrotron
spectrum ($f_\e^{syn}$), and the variability timescale ($t_{v,min}$),
and two unknowns, $\dD$ and $B$.  We use spectral modeling to
constrain these two unknowns.  Note the quantity $t_{v,min}\dD^3 B$ is
nearly, but not quite, a constant due to the appearance of $\dD$ in
the integrand and the limits.

\subsection{SSC Emission with Exact Synchrotron Expression}
\label{SSCexact}

\citet{crusius86} have derived an expression for the synchrotron emissivity
from isotropic electrons in a randomly-oriented magnetic field \citep[see also][]{ghisellini88},   
\begin{equation}
\e^{\prime} J_{syn}^{\prime}(\ep) = \frac{\sqrt{3}\e^{\prime}e^3 B}{h}
                \int^{\infty}_1 d\gp\ N_e^\prime(\gp)\ R(x)
\end{equation}
where $e$ is the fundamental charge, $h$ is Planck's constant, 
$$
x = \frac{4\pi \ep m_e^2 c^3}{3eBh\g^{\prime 2}}\ ,
$$
$$
R(x) = \frac{x}{2}\int_0^\pi d\theta\ \sin\theta\ 
\int^\infty_{x/\sin\theta} dt\ K_{5/3}(t), 
$$
and $K_{5/3}(t)$ is the modified Bessel function of the second kind 
of order 5/3.  The function $R(x)$ can be approximated as follows:
\begin{equation}
\log(R) = A_0 + A_1 y + A_2 y^2 + A_3 y^3 + A_4 y^4 + A_5 y^5
\end{equation}
where $y = \log(x)$ and the values of the coefficients are given in
Table \ref{Rtable}.  These approximations are accurate to $\sim 1$\% in
the range $10^{-2} < x < 10^1$; outside this range, the asymptotic
expressions from \citet{crusius86}, 
\begin{equation}
R(x) = \left\{ \begin{array}{ll}
	1.80842\ x^{1/3} & x \ll 1 \\
	\frac{\pi}{2}e^{-x}\left[ 1 - \frac{99}{162x}\right] & x \gg 1 \\
	\end{array}
	\right. \ ,
\end{equation}
can be used.  They are accurate to better than 5\% 
outside the range $10^{-2} < x < 10^1$. 

The synchrotron flux is then given by
\begin{equation}
\label{fesynexact}
f_\e^{syn} = \frac{\dD^4 \e^{\prime} J_{syn}^{\prime}(\ep)}{4\pi d_L^2}
 = \frac{\sqrt{3}\dD^4 \e^{\prime}e^3 B}{4\pi h d_L^2}
   \int^{\infty}_1 d\gp\ N_e^\prime(\gp)\ R(x)\ .
\end{equation}
Using this and eqs. (\ref{uprime}) and (\ref{epsjssc}) one 
gets the SSC emissivity, 
\begin{equation}
\e_s^{\prime} J_{SSC}^{\prime}(\ep_s) = 
\frac{9\sT d_L^2 \e^{\prime 2}_s}{4\dD^4 R^{\prime 2}_b}
\int^{\infty}_0\ d\ep\ \frac{f_\e^{syn}}{\e^{\prime 3}}\ 
\int^{\g^\prime_{max}}_{\g^\prime_{min}} d\g^{\prime}\ 
\frac{N^{\prime}_e(\gp)}{\g^{\prime 2}} F_C(q,\Gamma)\ ,
\end{equation}
and the observed flux (using eq.\ [\ref{feSSC}]), 
\begin{equation}
\label{feSSCexact}
f_{\e_s}^{SSC} = \frac{9}{16} \frac{ (1+z)^2\sT \e^{\prime 2}_s}
{\pi\dD^2 c^2 t_{v,min}^2 } \int^\infty_0\ d\ep\ 
\frac{f_\e^{syn}}{\e^{\prime 3}}\ \int^{\gp_{max}}_{\gp_{min}}\ d\gp\ 
\frac{N^{\prime}_e(\gp)}{\g^{\prime 2}} F_C(q,\Gamma)\ .
\end{equation}

\subsection{Comparison of Exact Synchrotron Expression with 
$\delta$-approximation}
\label{comparison}

In Fig.\ \ref{synchcompare} we compare the exact synchrotron expression
for synchrotron radiation, eq.\ (\ref{fesynexact}), with the
$\delta$-approximation, eq.\ (\ref{fes}).  In this calculation, we use
\begin{equation}
N_e^\prime(\gp) = K_e\ \g^{\prime -p}\ \exp\left(\frac{-\gp}{\gp_c}\right)\ H( \gp - \gp_1)
\end{equation}
for the electron distribution, with $K_e = 10^{49}$, $\gp_1 = 10^2$,
$\gp_c = 10^{3}$, $z = 0.116$, and various values of $p$.  We use $\dD
= 100$, $B = 10$ mG, and $t_{v,min} = 300$ s.  This figure shows that
the approximation is quite accurate near the center of the spectrum,
but loses accuracy at low and high frequencies.  At the
high frequencies, in particular, the improved accuracy of the full
expression is needed to accurately fit the VHE $\g$-ray data.

\subsection{SSC in the Thomson Regime}
\label{thomsoncompare}

Here we compare results using the Thomson cross section with the full
Compton cross section derived above.  Representing the synchrotron
spectrum as a monochromatic radiation field with comoving energy
density $u^\prime_{syn}$, the SSC flux can be approximated in the
Thomson regime, analogous to eq.\ (\ref{fes}), by the expression
\begin{equation}
f_{\e_s}^{SSC,T} \cong {\dD^4\over 6\pi d_L^2} \; c\sigma_{\rm T} 
u^\prime_{syn}
\gamma_T^{\prime 3} N_e^\prime (\gp_T)\ ,
\end{equation}
where 
\begin{equation}
\label{gamma_pT}
\gp_T = \sqrt{ \frac{(1+z)\e_s}{\dD\ep} }\ .
\end{equation}
Inserting eqs. (\ref{uprime}) and (\ref{Neprimegps}) into this, 
one gets
\begin{equation}
label{festhom}
f_{\e_s}^{SSC,T}  \cong \frac{ 24\pi}{c^3}\ 
	      \frac{ (1 + z)^2d_L^2
	      f_\e^{syn}(\e_{syn}^{pk}) f_\e^{syn}(\e_{syn}) }
	      {(t_{v,min} B \dD^3)^2}\ ,
\end{equation}

where
\begin{equation}
\label{esyn}
\e_{syn} = \frac{\e_s\e_B\dD}{\e_{syn}^{pk}(1+z) }\ .
\end{equation}
Here, if $f_{e_s}^{SSC}$ and $f_\e^{syn}$ are given from observations,
and the redshift (and hence the luminosity distance) of the object is
known, then the quantity $t_{v,min} B \dD^3$ is a constant in the
Thomson regime.  Due to the appearance of $\dD$ in the integrals of
eq.\ (\ref{feSSC}) or (\ref{feSSCexact}), this dependence does not
strictly hold when the full Compton cross section is used, but
describes the general behavior.  The constraint 
$t_{v,min} B \dD^3 = \textrm{constant}$ was derived
previously by \citet{tavecchio98}.  

In order to treat SSC emission in the Thomson regime in the
formulation of \S\ \ref{SSCdelta} or \S\ \ref{SSCexact}, one need only
replace eq.\ (\ref{jesC}) by
\begin{equation}
\label{Thomkernel}
F_T(q_T) = \frac{2}{3} \left(1\ -\ q_T \right)
H\left(q_T;\frac{1}{4\g^{\prime 2}},1 \right) \ ,
\end{equation} 
where $q_T = \ep_s / (4 \g^{\prime 2}\ep)$.  A comparison between the
Compton and Thomson calculations for a particular model with a
power-law electron distribution with an exponential cut off is shown in
Fig.\ \ref{thomson}.  As can be seen, the full Klein-Nishina expression
gives substantially different results, and would result in
significantly different parameter fits.  This demonstrates that the
full Klein-Nishina expression is necessary to do accurate spectral
modeling at $\g$-ray energies.  

\subsection{$\gamma\gamma$ Photoabsorption}
\label{photoabs}

\subsubsection{Exact Internal Photoabsorption}

Here we calculate the $\g\g$ absorption optical depth, $\tau_{\g\g}$,
due to interactions with the internal synchrotron radiation field.
This will, naturally, be a function of the synchrotron spectrum,
$f^{syn}_\e$.  Absorption will modify the high energy SSC spectrum by
the factor 
$$
\frac{ 1\ -\ e^{-\tau_{\g\g}} }{\tau_{\g\g}} ,
$$ 
and the absorbed $\gamma$-rays will be reinjected to form a second
injection component of high-energy leptons.  In this paper, we neglect
the additional cascade $\gamma$-rays formed by pair reinjection, which
is a good assumption when the absorbed energy is a small fraction of
the total $\gamma$-ray energy.

The photoabsorption optical depth for a $\gamma$-ray photon with
energy $\e_1$ in a radiation field with spectral photon density $n(\ep
,\mup ;r^\prime)$ is \citep{gould67,brown73}
\begin{eqnarray}
\tau_{\gamma\gamma}(\ep_1) =
\int_{r_1^\prime}^{r_2^\prime} dr^\prime \int_{-1}^1 d\mup (1-\mup)\ 
\int_{2/\ep_1(1-\mup )}^\infty d\ep\; \sigma_{\gamma\gamma}
[\ep\ep_1(1-\mup)]n^\prime(\ep ,
\mup ;r^\prime)\ .
\end{eqnarray}
For a uniform isotropic radiation field in the comoving frame, 
$n_{rad}(\ep,\mup ;r^\prime) \approx n_{rad}(\ep)/2$, so that
\begin{equation}
\tau_{\gamma\gamma}(\ep_1)
\cong R_b^\prime \int_0^\infty d\ep \; \sigma_{\gamma\gamma }
(\ep,\ep_1)n_{rad}^\prime(\ep )\ .
\label{taugg}
\end{equation}
Inserting the form of the absorption cross-section, one gets
\begin{eqnarray}
\label{taugg2}
\tau_{\g\g}(\ep_1) = \frac{R_b^\prime \pi r_e^2}{\e_1^{\prime 2} }
\int^\infty_{1/\ep_1} d\ep\ n_{rad}^\prime(\ep)\ \bar{\phi}(s_0)\ 
\end{eqnarray}
where $s_0=\ep\ep_1$, 
\begin{eqnarray}
\bar{\phi}(s_0) = \frac{1+\beta_0^2}{1-\beta_0^2}\ln w_0 - 
\beta_0^2\ln w_0 - \frac{4\beta_0}{1 - \beta_0^2}
\\ \nonumber 
+ 2\beta_0 + 4\ln w_0 \ln(1+w_0) - 4L(w_0)\ ,
\end{eqnarray}
$\beta_0^2 = 1 - 1/s_0$, $w_0=(1+\beta_0)/(1-\beta_0)$, and 
\begin{eqnarray}
L(w_0) = \int^{w_0}_1 dw\ w^{-1}\ln(1+w)\ .
\end{eqnarray}
Substituting the internal synchrotron radiation field, eq.\
(\ref{nprimesynep}), for $n_{rad}^\prime(\ep)$ in eq.\ (\ref{taugg2})
one gets
\begin{eqnarray}
\label{taugg3}
\tau_{\g\g}(\ep_1) = \frac{9d_L^2\sT(1+z)}
{8m_ec^6t_{v,min}\dD^5 \e^{\prime 2}_1}
\int^\infty_{1/\ep_1}\frac{d\ep}{\e^{\prime 2}} f_\e^{syn}\bar{\phi}(s_0)\ ,
\end{eqnarray}
where $\ep$ and $\e$ are related by $\ep = (1+z)\e/\dD$ 
(see eq.\ [\ref{gamma_ps}]).

\subsubsection{$\delta$-Approximation for Internal Photoabsorption}

An accurate approximation for internal $\g\g$ opacity is given by the 
$\delta$-function approximation 
\begin{equation}
\sigma_{\g\g}(\ep,\ep_1) \approx \frac{1}{3}\sT\ep
\delta\left(\ep-\frac{2}{\ep_1}\right)
\end{equation}
for the cross section \citep{zdziarski85}.
Eq.\ (\ref{taugg}) becomes
\begin{equation}
\tau_{\g\g}(\ep_1) = \frac{ \sT d_L^2\ep_1 }{ 2 m_ec^3R^\prime_b\dD^4 }
f^{syn}_{\bar\e}\ 
\end{equation}
where 
$$
\bar\e=\frac{2\dD^2}{(1+z)^2\e_1}\ ;
$$
or 
\begin{equation}
\label{taugg4}
\tau_{\g\g}(\e_1) \cong \frac{ \sT d_L^2 }{ m_ec^2 t_{v,min}\dD^4 }\ 
{ f^{syn}_{\bar\e}\over\bar\e } \ = 
\frac{ (1+z)^2\sT d_L^2 }{ 2 m_ec^4 t_{v,min}\dD^6 }\ 
\e_1 f^{syn}_{\bar\e}\ 
\end{equation}
This expression can be used to derive asymptotes
and determine the importance of internal $\g\g$ absorption.

Using this expression, and the fact that the blob must
be transparent to $\g$-rays, it is possible to derive a lower limit on
the Doppler factor.  Requiring $\tau_{\g\g} < 1$ and the assumption
that the synchrotron flux is well-represented by a power-law of index
$a$ ($f^{syn}_{\e} \propto \e^a$) leads to
\begin{equation}
\label{dDlimit}
\dD > \left[ \frac{ 2^{a-1} (1+z)^{2-2a} \sigma_T d_L^2}
          {m_e c^4 t_{v,min} } \e_1 f_{\e_1^{-1}}^{syn} 
      \right]^{\frac{1}{6-2a}}
\end{equation} 
\citep{dondi95}.

\subsubsection{Photoabsorption by the IBL}

Absorption by pair production will also occur due to interactions with
the IBL (modifying the observed spectrum by a factor of
$e^{-\tau_{\g\g}}$). In fact, IBL absorption is found to dominate the
internal $\g\g$ absorption for the blazars considered here where we
require large Doppler factors to produce hard-spectra multi-TeV
emission.  Formulas for absorption due to the IBL can be found in
\citet[][hereafter S06]{stecker06a}, \citet{stecker06b}, and
\citet{stecker07}.  However, it has recently been pointed out that
their approximations may overestimate the IBL \citep[][hereafter
D07]{dermer07cr}, which is more consistent with the calculations of
\citet[][hereafter P05]{primack05}; thus we use both the S06 and D07
formulations for $\tau_{\g\g}$ from the IBL.  {A comparison of the
opacity from the IBL formulations of S06, D07, and P05 for \object{PKS
2155--304} can be found in Fig.\ \ref{IBLcompare}.

\subsection{Jet Power and Constraints on $B$}
\label{jetL}

The synchrotron emission implies a minimum total (particle and field) 
jet power, for a given 
Doppler factor, variability timescale, and magnetic field.  The 
magnetic field which minimizes this jet power will be called $B_{min}$.  
Modeling the SSC component implies the departure of the 
magnetic field and jet power from these values. 

We use the
$\delta$-approximation to deduce the nonthermal electron spectrum
from the synchrotron spectrum, from which the total nonthermal electron 
energy and nonthermal jet electron power can be derived.
Adding the hadron and magnetic field energy gives a jet power that 
can be compared with the Eddington luminosity.  
The comoving energy in the
magnetic field is given by
\begin{equation}
\label{Benergy}
W^\prime_B = V_b^\prime U_B = \frac{R^{\prime 3}_bB^2}{6}\;,
\end{equation}
while the total comoving energy in the electrons is given by
\begin{equation}
\label{elenergy}
W^\prime_e = m_ec^2\int^{\gp_2}_{\gp_1} d\g^\prime_s\ \g^\prime_s\  
N^\prime_e(\g^\prime_s)\ .
\end{equation}
Substituting the electron distribution, $N^\prime_e(\g^\prime_s)$ from
eq.\ (\ref{Neprimegps}) into eq.\ (\ref{elenergy}) and recalling eq.\
(\ref{gamma_ps}), $ \g_s^{\prime 2} = \e(1+z)/(\dD\e_B)\ , $ one
obtains
\begin{equation}
W^\prime_e = m_e c^2 \frac{6\pi d_L^2}{c\sigma_T\e_B^2U_{Bcr}\dD^4}\ 
\frac{1}{2}\ \sqrt{ \frac{\dD\e_B}{1+z} }\ I_{syn}\;,
\end{equation}
where 
$$
I_{syn} = \int_{\dD\e_B\gp_{1}/(1+z)}^{\dD\e_B\gp_{2}/(1+z)} d\e \frac{f_\e^{syn}}{\e^{3/2}}\ 
$$
and $U_{Bcr} = B_{cr}^2/(8\pi)$.  Thermal protons or protons
co-accelerated with electrons will also contribute to the total
particle energy. The ratio of the total particle to electron energies
is given by $\xi \equiv W^\prime_{par} / W_e^\prime$, so that the
energy in all of the particles (electrons and protons) is
\begin{eqnarray}
\label{W_par}
W^\prime_{par}  \; = \; \xi W_e^\prime 
 \; = \; \frac{ 3\pi d_L^2\xi m_ec^2}
      { c\sigma_T\e_B^{3/2}(1+z)^{1/2} U_{Bcr}\dD^{7/2} }
      \ I_{syn}\;.
\end{eqnarray}
In our calculations, we set $\xi = 10$.

The total jet power in the stationary frame 
(i.e., the frame of the galaxy), is given by
\begin{equation}
\label{jetpower}
P_j = 2 \pi R_b^{\prime 2}\beta\G^2c \frac{W_{tot}^\prime}{V_b^\prime}
\end{equation}
\citep{celotti93,celotti07}, where $W^\prime_{tot} = W^\prime_{par} +
W^\prime_B$, and the factor of 2 is due to the assumption that the
black hole powers a two-sided jet.  This formulation gives parameters
that minimize black hole jet power \citep{dermer04}, though without
taking into account whether these parameters provide a good spectral
fit.  The magnetic field that minimizes the jet power $P_j$is obtained by
solving $|{dP_j/d\epsilon_B}|_{\epsilon_B = \e_{B,min}} = 0$ for $\e_B
= B/B_{cr}$.  Doing this gives
\begin{equation}
\e_{B,min}  = \left[ \left( \frac{3}{2} \right)^3\ 
\frac{ \xi m_ec^2d_L^2 (1+z)^{5/2} }
     { 2t_{v,min}^3c^4 \sT U_{Bcr}^2 \dD^{13/2} }\ I_{syn} \right]^{2/7}\ .
\label{ebmin}
\end{equation}
Smaller or larger magnetic field values than $B_{cr}\e_{B,min}$ are possible, 
but only for more powerful jets.  The minimum $P_j$ becomes
\begin{equation}
P_{j,min} = \frac{14}{3}\ 
            \pi \;R_b^{\prime 2}\e^2_{B,min}\beta\Gamma^2 U_{Bcr}\;= \frac{14}{3}\ 
            \pi c^3\left[ \frac{\dD \G t_{v,min}\e_{B,min}}{1+z}\right]^2
            \beta U_{Bcr}\;,
\end{equation}
which makes the ratio of the jet power to the minimum jet power 
\begin{equation}
\frac{P_j}{P_{j,min}} = 
\frac{3}{7} \left[  \zeta_B^2 + 
\frac{4}{3} \zeta_B^{-3/2} \right]\;,
\end{equation}
where $\zeta_B = \e_B/\e_{B,min}$.

It is also useful to compare the jet power with the total 
luminosity, which is given by
\begin{equation}
L_{tot} \cong \frac{2\pi d_L^2}{\Gamma^2}\ 
\int_0^\infty  d\e\ \frac{f_{\e}}{\e} ,
\end{equation}
including the beaming factor $1/2\Gamma^2$ for a two-sided jet. The radiative
efficiency is given by $L_{tot}/P_j$.

\subsection{Fitting Technique}
\label{fittech}

We calculate the synchrotron spectrum with eq.\ (\ref{fesynexact})
using a model for the comoving electron spectrum, and calculate the
SSC spectrum with eq.\ (\ref{feSSCexact}) taking into account internal
$\g\g$ absorption, eq.\ (\ref{taugg2}), and IBL $\g\g$ absorption.  The
synchrotron and SSC components are fit separately and iteratively in
the following manner (see Fig.\ \ref{fitdiagram}): Starting values are
chosen for the electron distribution parameters (see \S\ \ref{edist}),
$\dD$ and $B$.  The synchrotron spectrum calculated with
eq.\ (\ref{fesynexact}) is fit to the low energy (in this case, {\em
Swift}) data points by varying the electron distribution but keeping
$\dD$ and $B$ constant, using a $\chi^2$ minimization technique
\citep[i.e., the Method of Steepest Descent; e.g., ][]{press92}.  Once
the electron distribution is obtained from this fit, the high energy
data are fit to the SSC component, calculated from
eq.\ (\ref{feSSCexact}) and $\chi^2$ is calculated.  Then the
parameters $\dD$ and $B$ are varied, the synchrotron spectrum is again
fit to the low energy data to obtain the electron distribution, and
another $\chi^2$ is calculated from the high energy data.  This
process is repeated until the parameters ($\dD$ and $B$) that minimize
$\chi^2$ are found.  The constants of the problem are the
multiwavelength spectral data, the source redshift $z$, variability
timescale $t_{v,min}$, and the lower Lorentz factor $\gp_1$ of the
electron distribution.

In order to account for systematic errors in the VHE data, 
which usually dominates the measurement errors,
we performed fits by multiplying the VHE data by a factor of 
$N_{sys} (\nu/\nu_{mid})^{a_{sys}}$ where $\nu_{mid}$ is the geometric 
mean of the VHE data, and $N_{sys}$ and $a_{sys}$ fit parameters that 
were allowed to vary within the systematic error range.  $N_{sys}$ 
accounts for the normalization error, and $a_{sys}$ accounts for the 
spectral index error; for no systematic errors, $N_{sys}=1$ and 
$a_{sys}=0$.  

\section{APPLICATION TO PKS 2155--304}
\label{pks2155}

The XBL \object{PKS 2155--304}, an EGRET source
\citep{hartman99} and one of the brightest blazars at TeV energies,
has been the subject of several multiwavelength campaigns
\citep[e.g.,][]{zhang01, osterman07,sakamoto07}.  The redshift of
\object{PKS 2155--304} is $z=0.116$ \citep{falomo93,sbarufatti06b},
giving a luminosity distance of $d_L=540$ Mpc in a cosmology where
$H_0=70$ km s$^{-1}$ Mpc$^{-1}$, $\Omega_M=0.3$, and
$\Omega_\Lambda=0.7$.

In July and August of 2006, the source underwent several extremely
bright flares which were detected by {\em HESS} \citep{aharonian07_2155}
and followed up by {\em Swift} \citep{benbow06,foschini07}.  The
measured variability timescales of the flares were as short as a few
minutes with {\em HESS}.  {\em Swift}'s observing schedule did not
allow it to probe such small timescales; however, {\em BeppoSAX}
observations from 1996 to 1999 show X-ray variability on scales of
$\sim 1$ hour \citep{zhang02}.  Preliminary analysis of {\em Chandra}
data taken simultaneously with the {\em HESS} observations show them
to be strongly correlated, and thus have X-ray variability timescales
of a few minutes as well, although ground-based 
observations show the optical to be uncorrelated with the 
X-rays and VHE $\g$-rays \citep{costamante07}.  Rapid variability 
does not seem to be intrinsic to only \object{PKS 2155--304}, 
as variability on timescales of a few minutes at TeV energies has
been observed from \object{Mkn 501} as well \citep{albert07}.  
The lack of optical correlation makes it unlikely that the 
optical emission came from the blob where the flare originated, and   
hence we consider the {\em Swift} optical data an upper limit.  

The data used in our analysis are not simultaneous; the {\em Swift}
ultraviolet/optical telescope (UVOT) and X-ray telescope (XRT) data
are from 30 July, when it first observed the flares, while the {\em
HESS} data are from 28 July.  Detailed simultaneous {\em HESS} data
have not yet been made available, although \object{PKS 2155--304} was
detected by {\em HESS} on 30 July simultaneously with the {\em Swift}
detection \citep{foschini07}.

Using eqs.\ (\ref{fesynexact}) and (\ref{feSSCexact}), and taking
taking into account $\g\g$ absorption by internal jet radiation (eq.\
[\ref{taugg2}]) and the IBL, we simulate the broadband emission during
one of these extremely bright flares.  We fit the fully reduced
{\em Swift} and {\em HESS} data, with the {\em Swift} data corrected
by \citet{foschini07} for a Galactic
column density of $N_H = 1.36\times10^{20}$
cm$^{-2}$.

\subsection{The Electron distribution}
\label{edist}

The electron distribution is assumed to be of the form
\begin{eqnarray}
\label{elec_dist}
N_e(\gp) = K_e\Biggr[ \left(\frac{\gp}{\gp_{break}} \right)^{-p} H(\gp_{break} - 
\gp) + 
\\ \nonumber
\left(\frac{\gp}{\gp_{break}}\right)^{-(p+1)} H(\gp - \gp_{break}) \Biggr]
H(\gp;\gp_1,\gp_2)\;.
\end{eqnarray}
As long as $\gp_1\lesssim \gp_{min}$ (eq.\ [\ref{gpmin}]) 
and $\gp_2\gg \gp_c$, the values
of $\gp_1$ and $\gp_2$ do not affect our fits.  This leaves us with
four fitting parameters to define the electron spectrum, 
$K_e$, $p$, and $\gp_{break}$.  This spectrum is physically
motivated, in that radiative cooling is expected to modify the electron
power-law index by 1.  For all of out fits to \object{PKS 2155--304},
we found $p = 2.7$.  

The electron distribution fitting parameters are not unique, and
several sets of parameters give equally good fits to a given
synchrotron spectrum.  The fit of the SSC spectrum to the $\gamma$-ray
data and the parameters $\dD$ and $t_{v,min}$ derived in the fitting
routine are, however, essentially independent of electron distribution
parameters, as long as they provide a good fit to the synchrotron
component.

\subsection{The SSC Spectrum}
\label{SSCpks2155}

The {\em Swift} and {\em HESS} data from \object{PKS 2155--304} were
fit using the technique described in \S\ \ref{fittech}.  Results of
the fits can be seen in Table \ref{parameters} and in
Figs. \ref{SEDtv300} and \ref{SED_dermer}.  The parameters $t_{v,min}$
and $\gp_{1}$ were kept constant during the fits, and the angle
along the line of sight was assumed to be small, so that 
$\Gamma \approx \dD$ for the purpose of calculating the jet power.
When we did vary $\gp_1$, we found that this had no significant effect
on our fits, as long as it was sufficiently low.  All the models
presented here have $\gp_1=10^3$.  Based on the systematic errors
reported in \citet{aharonian07_2155}, $N_{sys}$ was allowed to vary
between 0.8 and 1.2, and $a_{sys}$ was allowed to vary between $-0.1$
and $+0.1$.  This leads to the error bars on $\dD$ and $B$ in Table
\ref{parameters}, which are systematic error bars.  

The fit parameters are strongly dependent on $t_{v,min}$.  As
$t_{v,min}$ increases, the best fit $\dD$ decreases.  This is mainly
due to the relationship defining the SSC spectrum in eq.\
(\ref{feSSC2}), which remains roughly constant provided that the
quantity $t_{v,min}\dD^3 B$ remains constant, which is true for
scattering in the Thomson regime, as shown below.  

Based on the {\em Swift} and {\em HESS} data, eq.\ (\ref{dDlimit}) 
gives the constraint
\begin{equation}
\label{dopplerpks}
\dD \ga 48 \left( \frac{300\ \sec}{t_{v,min}}\right)^{0.175}\ .
\end{equation}
Thus, the short variability in \object{PKS 2155-304} \citep[$\approx
300$ sec,][]{aharonian07_2155} limits the Doppler factor to quite high
values \citep[see also][]{begelman08}.
Another constraint can be found from eq.\ (\ref{esyn}) by noting 
that the peak energy for the SSC component must be $> 0.2$ TeV 
($\e_s > 4 \times 10^5$).  This leads to
\begin{equation}
\left(\frac{B}{1\ \textrm{G}}\right)\dD \ga 7
\end{equation}
\citep{tavecchio98,bednarek97,bednarek99}.

The best fit magnetic field values implied by spectral modeling are a
few to $\approx 100\ $ mG, which represent a small
fraction of the magnetic field, eq.\ (\ref{ebmin}), that minimizes the
jet power.  The absolute jet powers for a two-sided jet derived from
these fits are $\approx 10^{46}$ -- $10^{47}$ ergs s$^{-1}$, representing
a very large fraction of the Eddington luminosity for a $10^9 M_\odot$
black hole ($L_{Edd} = 1.3 \times 10^{47}$ erg s$^{-1}$). Values this
large would cast doubt on the underlying one-zone synchrotron/SSC and
IBL model.  Also note that the small values of $\zeta_B$ indicates
that $W^\prime_{par} \gg W^\prime_B$; thus, $P_j \propto \xi$, and if
$\xi$ is any larger, the jet power will be larger by the same factor.
Because the jet power is strongly particle-dominated,
it is very dependent on the value of $\gp_1$, which is not strongly
constrained by our models.  However, we take $\gp_1 = 10^3$, a rather
large value.  It is unlikely to be larger, and thus, the jet power is
unlikely to be lower, although this part of the electron distribution
is not well-constrained by observations due to the lack of
correlated optical variability.  Our choice of electron spectral
index assumes the blob is primarily cooled by synchrotron or SSC
emission, which may not be the case (see below).  

Fits were performed with the IBL models of (in order of
decreasing intensity) S06, D07, and P05.  The lower-intensity IBLs do
lower the Doppler factors and jet powers considerably.  This may be an
arguement for the lower IBLs, which has also been suggested from
observations of TeV observations of 1ES 1101--232 \citep{aharonian06}
and 1ES 0229+200 \citep{aharonian07_0229}.  However, the jet powers
are still quite large compared to the probable Eddington limit.  The
lowest Doppler factor we were able to obtain, with the lowest IBL and
longest $t_{v,min}$ was $\dD=58$.  This Doppler factor is
still considerably larger than values measured with the VLBA from
{\object{PKS 2155--304} on the parsec scale \citep[$
\beta_{obs} \sim 4$;][]{piner04}.  The $\gamma$-ray
emission may, however, be formed on size scales that are tens to
hundreds of times smaller.  Note that the models differ significantly
in the frequency region that {\em GLAST} will observe (i.e., $\sim 10$
MeV---100 GeV); thus, {\em GLAST} will be quite useful for
distinguishing between models.

\subsection{Temporal Variability}
\label{tempvar}

Temporal variability can result from  
acceleration, adiabatic expansion, and radiative cooling.
Nonthermal electrons that cause blazar flares are thought to be
accelerated to high energies by a Fermi acceleration mechanism,
possibly caused by internal shocks in the collision of irregularities
in the relativistic jetted wind.  
The particle acceleration timescale in the comoving frame for
a Fermi mechanism is
\begin{equation}
t_{acc}^\prime = N_a\ \frac{\gp}{\nu_B}
\label{tacc}
\end{equation}
where $N_a\gtrsim 1$ is the number of gyrations an electron 
makes while doubling its energy, and 
$$
\nu_B = \frac{eB}{2\pi m_ec}\ 
$$
is the Larmor frequency \citep[e.g.,][]{gaisser90}.

As the electrons radiate synchrotron and Compton-scattered radiation, 
they lose energy.  The electron energy-loss rate, or cooling rate,
in the comoving frame from synchrotron
radiation is
\begin{equation}
-\dot{\gamma^\prime}_{syn} = \frac{4}{3}c\sT u_B\gamma^{\prime 2}.
\end{equation}
The SSC cooling rate using the full Klein-Nishina cross section
\citep{jones68,boett97} is
\begin{equation}
-\dot{\g^\prime}_{SSC} = \frac{3 \sT}{8m_ec}\int^{\infty}_{0}d\ep 
	         \frac{u^\prime(\ep )}{\e^{\prime 2}}\ G(\g^{\prime}\ep)
\end{equation}
where 
$$
G(E) = \frac{8}{3}E\frac{1+5E}{(1+4E)^2}\ -\ 
	\frac{4E}{1+4E}\left(\frac{2}{3}+\frac{1}{2E}+\frac{1}{8E^2}\right)
$$
$$
	+\ \ln(1+4E)\left( 1+\frac{3}{E}+\frac{3}{4}\frac{1}{E^2} + 
	\frac{\ln[1+4E]}{2E} - \frac{\ln[4E]}{E} \right)
$$
\begin{equation}
-\ \frac{5}{2}\frac{1}{E}\ +\ \frac{1}{E}\sum^{\infty}_{n=1}
\frac{(1+4E)^{-n}}{n^2}\ -\ \frac{\pi^2}{6E}\ -\ 2\;,
\end{equation}
and $u^\prime (\ep) $ is the synchrotron spectral energy density, given by 
eq.\ (\ref{uprime}).
With these expressions one can determine the cooling 
timescales in the observer's frame, given by
\begin{equation}
t_{cool}\ =\ \frac{1+z}{\dD}\ t^\prime_{cool}\ =\ \frac{1+z}{\dD}\ 
\vline\frac{\g^\prime}{\dot{\g}^\prime}\vline\ .
\end{equation}

The acceleration and cooling timescales can be seen in Fig.\
\ref{cooling} for Model 3 with $t_{v,min} =$ 300 s and $N_a = 10$ and
1000; results for other models are similar.  The SSC cooling timescale
starts to deviate from the Thomson regime behavior at $\gp \ga 2 \times10^5$
due to Klein-Nishina effects, which is not seen on this plot.
Overplotted on this graph is the electron spectrum used to fit the
\object{PKS 2155--304} data, which cuts off at $\gp =2\times
10^5$.  In \object{PKS 2155--304}, the cutoff in the synchrotron
spectrum at $\nu \approx 10^{18}$ Hz means that electrons are not
accelerated to extremely large Lorentz factors. If the cutoff in the
synchrotron spectrum and therefore the electron spectrum is attributed
to cooling that prevents acceleration to higher energies, then $N_a$
must be very large, $\gtrsim 10^5$.  This could occur if there are
more severe limitations to electron acceleration than implied by the
simple expression given by eq.\ (\ref{tacc}). See \S\ \ref{discuss}
for further discussion.

\section{APPLICATION TO MKN 421}

The XBL \object{Mkn 421}---the first blazar seen at TeV
energies---has been the target of many multiwavelength campaigns
\citep[e.g.,][]{macomb95, fossati04, fossati08}.  The campaign of
March 2001 was one of the longest and most complete \citep{fossati08},
in which it was observed by {\em RXTE}, {\em Whipple}, {\em HEGRA},
and the Mt.\ Hopkins 48\arcsec\ telescope nearly continuously for
seven days.  During this campaign on 19 March 2001, an extremely
bright flare was observed with all three instruments.  In this
section, we model this flare with our SSC methodology.  
\object{Mkn 421} has shown variability on timescales down to $\sim
1000$ s \citep[e.g.,][]{aharonian02}.  This XBL has a redshift of
$z=0.03$ \citep{ulrich75, ulrich78} giving it a luminosity distance of
$d_L=130$ Mpc using the cosmological parameters mentioned in \S\
\ref{pks2155}.

Fits to the March 2001 flare for different $t_{v,min}$
and IBL formulations are shown in Table \ref{Mknparameters} and Fig.\
\ref{Mkn421}.  We fit to the fully reduced data from
\citet{fossati08}.  The {\em RXTE} data was reduced assumed a Galactic
column density of $1.6\times10^{20}$ cm$^{-2}$, although this makes
little difference for {\em RXTE}'s bandpass.  The same electron
distribution as for \object{PKS 2155--304} was used, namely eq.\
(\ref{elec_dist}), except for \object{Mkn 421}, the best fit electron
power-law index was $p=2.2$.  Again, it is assumed $\dD \approx
\Gamma$.

As with \object{PKS 2155--304}, for a constant $t_{v,min}$, as $\dD$ 
increases, $B$ decreases.  The opacity constraint, eq.\ (\ref{dDlimit}), 
gives 
\begin{equation}
\dD \ga 29 \left( \frac{10^3\ \sec}{t_{v,min}} \right)^{0.2} \ .
\end{equation}
The S06 and D07 IBLs are nearly identical for $z=0.04$.  All IBLs
result in a low opacity of intergalactic space at such low $z$.
Absorption by the IBL plays a small, but non-negligable part for
low-$z$ VHE $\g$-ray sources.  However, the difference between the IBL
formulations is negligible, as the solutions with different IBLs are
all within the margin of error.

The model fits of \citet{fossati08} give $\dD=100$ and $B=1$ G for a
blob size of $R_b^\prime = 1\times10^{14}$ cm which is similar to our
fits, except with a larger $B$.  It should be noted that their models
do not fit the TeV data particularly well.  VLBA observations of
\object{Mkn 421} show only weak superluminal motion, with $\beta_{obs}
\sim 0.1$ on parsec scales \citep{piner05}.  As with \object{PKS
2155-304}, the jet must slow down considerably from the $\g$-ray
emitting region to the parsec-scale jet.  Also as before, $\zeta_B$ is
very similar in all fits.  

The cooling timescales for Model 13 for Mkn 421 can be seen in Fig.\
\ref{coolingMkn421}.  Here, the Klein-Nishina effects on the Compton
cooling timescale are visible at $\gp\sim 8\times10^4$, indicating
that accounting for Klein-Nishina effects is quite important for modeling
this flare.  As with \object{PKS 2155--304}, the cooling timescale is
longer than the variability timescale, which poses problems for the
SSC model.  

\section{DISCUSSION}
\label{discuss}

We have presented a new approach to modeling XBL radiation in a
synchrotron/SSC framework. We obtain the electron spectrum by fitting
a model nonthermal electron distribution to the low energy
radio/optical/X-ray data, assuming that it arises from nonthermal
synchrotron processes. This electron spectrum is then used to
calculate the SSC component and to fit the $\gamma$-ray data as a
function of a small set of parameters.  The number of free parameters,
namely the Doppler factor, magnetic field, and size scale of the
radiating region, is small enough that $\chi^2$ fits can be performed
to high quality multiwavelength data. We discuss the results given by
fitting the non-simultaneous \object{PKS 2155--304}
data (keeping in mind that unambiguous conclusions will require
simultaneous data sets), as well as simultaneous {\object Mkn 421}
data.

\subsection{The Rapid Flare in PKS 2155--304}

For the giant flare in \object{PKS 2155--304}, the rapid ($\sim 5$
min) variability implies a small emitting region.  This creates
problems for the one-zone SSC model for three main reasons: it
requires excessively large Doppler factors, jet powers, and cooling
timescales.  

For fits to the full optical/UV/X-ray spectra for \object{PKS
2155--304}, we find that very large Doppler factors (
$\dD \gtrsim 60$ for \object{PKS 2155--304}) are
necessary to explain high energy emission with the SSC mechanism
\citep[similar to the findings of] []{begelman08}.  However, radio
measurements of superluminal motion from \object{PKS 2155--304}
indicate that the jet Lorentz factor $\Gamma \sim 10$, at least on
parsec scales \citep{piner04}.  The cause of this discrepancy could be
that the flare radiation is produced very close to the black hole, and
that the blobs seen in the radio have slowed down before reaching
parsec scales.  However, a deceleration episode between the inner
regions near the black hole and the parsec-scale region would likely
produce observable high energy radiation distinct from the SSC
component \citep[e.g.,][]{georgan03,levinson07}.  If all 
blazars have such large Doppler factors, the number of 
blazars aligned with our line of sight becomes smaller 
than the number actually observed.  

The excessive jet powers required by our fits
to the full optical/UV/X-ray and $\gamma$-ray spectra are another 
problem for the one-zone SSC model.  The
jet powers  for \object{PKS 2155--304} range
from $\sim 10^{46}$ -- $4\times 10^{47}$ ergs s$^{-1}$ (see Table
\ref{parameters})}; these powers can be reduced
by a factor of 10 for a hadron-free pair jet, recalling that we
assumed $\xi = 10$.  The higher values exceed the Eddington
luminosity for a $\sim 10^9\ M_{\odot} $ black hole by a factor of a
few.  Even if the black hole exceeded $10^9 M_\odot$, it is unlikely
that a black hole with a jet would be so radiatively efficient. Large
Doppler factors in synchrotron/SSC modeling were also found in
analyses of \object{Mkn 501} by \citet{krawczy02}.

The jet powers of our fits are significantly larger than the jet
power, $6 \times 10^{43}$ ergs s$^{-1}$, found for the SSC fit to the
same \object{PKS 2155--304} data by \citet{ghisellini08}.  Their SSC
fit is closest to our Model 5, which underfit the optical {\em
Swift} data as well.  The electron jet power in Model 5 is a
factor of $\sim 40$ times larger than the SSC
model of \citet{ghisellini08}, when taking into account the fact that
they calculated it for a one-sided jet.  The main reason for this is
that their synchrotron component underfits the optical data by a
larger amount (also note that they did not use the full Compton
cross-section, which can significantly affect derived parameter values
and jet powers; see \S\ \ref{thomsoncompare}).  Even though the
bolometric power at optical energies is only a factor of $\sim 3$
greater than the X-ray power, the total jet power in a synchrotron/SSC
model that fits the optical as well as the X-rays is much greater than
the power to fit the X-ray and TeV radiation because of the many
low-energy radiatively inefficient electrons required to emit the
lower energy optical radiation in the spectrum.  It is also worth
reminding the reader that, as discussed in \S\ \ref{SSCpks2155}, the
low energy part of the electron distribution---and hence the jet
power--- is essentially unconstrained by optical observations.
However, it is difficult to imagine how $\gp_{1}$ could be much larger
than $10^{3}$, so that our jet powers could be considered lower
limits.  Our jet powers are also dependent on the assumption of
cooling dominated by synchrotron or SSC losses.  

Another important issue for the SSC model studied here is that one
would expect the timescale of variability at X-ray synchrotron and TeV
$\gamma$-ray energies generated by the same electrons would be
approximately equal\footnote{Actually, the variability timescale of
synchrotron X-rays should be {\em smaller} than at the corresponding
SSC $\g$-rays.  This is because photons with a wide range of energies
are scattered by electrons with a wide range of energies to create the
SSC emission, which tends to ``smear out'' the variability.  The
synchrotron variability, on the other hand, is due only to the
electron variability, not variability in a source photon field.  Thus,
the synchrotron's variability is not washed out as much as the SSC's.
}.  Equating eqs.\ (\ref{gamma_ps}) and (\ref{gamma_pT}), assuming
that the high-energy radiation is dominated by Thomson-scattered peak
synchrotron photons with energy $\e_{pk}$, implies that photons with
energy $\e_{s} \cong (1+z)\e_{pk}\e_{syn}/(\dD\e_B)$
(eq.\ [\ref{esyn}]) are produced by the same electrons that emit
synchrotron radiation with energies $\sim \e_{syn}$. The results of
Model 3 for \object{PKS 2155--304} in Table \ref{parameters} imply
that $\e_{s} \cong 3.3\times 10^9\e_{syn}$, taking $\e_{pk} \cong
10^{-4}$.  Electrons Compton scattering $\e_{pk}$ photons to make 1
TeV photons would therefore be radiating synchrotron photons at
$\approx 0.3$ keV with the same short variability timescale as
measured at TeV energies.  Because the SSC cooling timescale in the
Thomson regime scales as $\propto \g^{\prime -1}$, the higher energy
photons observed by {\em Swift}'s XRT should have an even shorter
variability timescale than observed at 1 TeV by {\em HESS}; however,
{\em Swift}'s limited observing schedule did not allow a detailed
variability study.  {\em Chandra} did observe the flare, and
preliminary analysis seems to indicate the X-rays are highly
correlated with the $\g$-rays observed by {\em HESS}
\citep{costamante07}, and strong X-ray/$\g$-ray correlation in 
other blazar flares has been observed before 
\citep[e.g.,][]{fossati08,sambruna00,tavecchio01}, 
but its variability timescale is still not
clear.  In any case, whatever the source of the variability is, the
timescale will be limited by the size scale of the emitting region.

The one-zone SSC model for the TeV flare in \object{PKS
2155--304} implies, from Fig.\ \ref{cooling}, that the radiative
cooling timescales are much longer than $t_{v,min}$.  Thus,
variability cannot be attributed to radiative cooling, but could
originate from adiabatic expansion; however, as mentioned above, the
lack of achromatic variability in \object{PKS 2155-304} makes this
seem unlikely.  Thus, the variability is not consistent with the 
one zone synchrotron/SSC model.  \citet{begelman08} use analytic
estimates to show that, with the brightness and temporal variability
observed in \object{PKS 2155--304}, $\Gamma \ga 50$ for the radiation
to avoid photoabsorption.  They also suggest that magnetic energy
density must dominate the jet power in order for the acceleration to
be efficient, which limits $\Gamma \la 40$; thus, the SSC model cannot
explain these flares (however, in our simulations, the jet power is
dominated by particle energy density [\S\ \ref{SSCpks2155}]).  They
suggest the flare is caused by Compton scattering of an external
radiation field that must be limited to below sub-mm wavelengths.
Using their inferred energy density for the external scattering
radiation, then unless the size scale of this region is $\la 3 \times
10^{16}$ cm, this radiation should be observable, which makes it more
likely to originate from an accretion disk.

Are there explanations that could resolve these issues with the rapid
TeV flare from \object{PKS 2155--304}?  We discuss three possibilites:
external Compton scattering, relaxing the assumption of homogeneity,
and lower IBL energy densities.

The SSC mechanism involving isotropic scattering is unable to
decelerate a jet blob from the excessive Doppler factors found from
our fits, to the Doppler factors required from radio observations
\citep{piner05}.  Jet deceleration would require another mechanism
such as external Compton scattering where jet electrons upscatter
photons of an external radiation field.  This would produce flares at
GeV energies.  These photons could originate from a slower-moving
sheath surrounding a faster jet spine
\citep{ghisellini05,tavecchio08}, or from an advanced portion of the
jet moving at a slower speed \citep{georgan03}.  The external
photons would also serve as a power and cooling source, which could
also resolve issues with excessive jet powers and cooling timescales.
A complete analysis would require an extension of the
synchrotron/SSC model to include external Compton scattering and
effects of the $\gamma\gamma$ opacity \citep{rei07} from the scattered
radiation field to account for the full broadband SED (C.\ D.\ Dermer,
J.\ D.\ Finke, H.\ Krug, \& M.\ B\"ottcher 2008, in preparation).  In
this case, the 100 GeV -- TeV radiation would still predominantly
arise from the SSC process because of strong Klein-Nishina effects on
external Compton components from scattered optical/UV radiation, but
the additional $\gamma\gamma$ opacity would have to be considered in
the analysis.

Another possible explanation for the extreme parameters is that
the homogeneous assumption for the blob is invalid. We have assumed
that the synchrotron radiation, which provides target photons for
Compton scattering, is emitted by the same nonthermal electrons,
whereas they could originate from a more extended region. The
correlated variability between X-rays and TeV $\gamma$-rays seen in
PKS 2155-304 \citep{foschini07,costamante07} indicate however that at
least the higher energy synchrotron photons are probably co-spatially
produced with the SSC radiation.  Even in a one-zone model, the
electrons could be strongly cooled before the synchrotron photons
uniformly fill the emission region.  Alternately, the entire blob
might not be optically thin to $\gamma\gamma$ attenuation and the
observed SSC emission may be from a smaller region than the
synchrotron emission.  Simulations with corrections to the one-zone
approximation are necessary to investigate these possibilities.  

Recent blazar observations may indicate an IBL energy density
only slightly above the lower limits implied from galaxy counts
\citep{aharonian06,aharonian07_0229}.  The lowest IBL considered here,
P05, gives significantly lower jet powers and Doppler factors for the
\object{PKS 2155--304} flare than the other models, which may be a
further arguement for a lower IBL.  The IBLs of D07 and P05 both give
jet powers below the Eddington limit for a $10^9\ M_\sun$ black hole.
However, the Doppler factors are still excessive.  

To determine how certain our fit from Model 5 is, we
plotted the 68\%, 95\%, and 99\% statistical uncertainty
contours.  Also plotted is the jet power as a function of the
parameters $\dD$ and $B$ for Model 5 in Fig.\ \ref{contour}.
The confidence contours trace a region that is roughly constant in
terms of jet power; inside the 68\% contour, the power varies between
a relatively small amount (between roughly $2\times10^{46}$ and
$10^{47}$ erg s$^{-1}$).  Also of note is that a relatively large
range of Doppler factors ($\dD \sim$ 80 -- 170 with 68\% confidence)
will give a satisfactory fit.  Even the lowest Doppler factors allowed
are quite high.  The Doppler factors are
too high for the eq.\ (\ref{dopplerpks}) constraint to play a part.
The lines of constant jet power follow a power law based on $B \propto
\dD^{-1.5}$, as one can see from eq.\ (\ref{jetpower}),
$$
P_j \propto R_b^{\prime -1}\dD^2 W^\prime_{par}\ \propto \dD W^\prime_{par}\ ,
$$
if $\dD \approx \Gamma$ and $W^\prime_e \gg W^\prime_B$.  Most of the
electron energy is located in the part of the electron spectrum below
the break.  Given a $\nu F_{\nu}$ spectral index of $a_1$
($f_e\propto\e^{a_1}$), and performing the integral in eq.\
(\ref{W_par}), $W_{par} = B^{a_1-2}\dD^{a_1-4}$.  For constant $P_j$,
this gives $B\propto\dD^{\frac{a_1-3}{2-a_1}}$.  For our fits to PKS
2155--304, $a_1=0.15$, and we recover $B\propto\dD^{-1.5}$.
Statistical uncertainty contours follow $B\propto\dD^{-2.4}$, rather
than the expected $B\propto\dD^{-3}$ (\S\ \ref{thomsoncompare}).  
This is due to the fact that scattering occurs in the Klein-Nishina 
regime.  For $\gp\ep\ga1/4$, to zeroth order the scattering cross-section goes 
as approximately $\sigma_{KN}\propto\e^{-1}$.  
At the peak energy, eq.\ (\ref{festhom}), using the Klein-Nishina 
cross section and eq.\ (\ref{esyn}) becomes
$$
f_{\e^{pk}_s}^{SSC}\propto \frac{\left[ f_\e^{syn}(\e^{pk}_{syn})\right]^2 }
	{B^2\dD^6\e_{syn}^{pk}} 
	\propto \frac{\left[ f_\e^{syn}(\e^{pk}_{syn})\right]^2 }
	{B^{5/2}\dD^{13/2}}\ ,
$$
which leads to $B\propto\dD^{-2.6}$.  This is very close to 
the $B\propto\dD^{-2.4}$, considering the accuracy of the 
approximation.
The internal $\g\g$ opacity, eq.\ (\ref{dopplerpks}), does not
significantly constrain the Doppler factor in Fig.\ \ref{contour}.  

\subsection{The March 2001 Flare in Mkn 421}

For \object{Mkn 421}, we find $\dD \ga 30$ for our
model fits.  VLBA observations of \object{Mkn 421} show only weak
superluminal motion, with $\Gamma \sim 2$ on parsec scales
\citep{piner05}.  Jet powers are one the order of $P_j\ga
3\times10^{45}$ erg s$^{-1}$, and for all of our models they do not
exceed the Eddington luminosity for a $10^9$ M$_\sun$ black hole.  The
parameters for this flare in \object{Mkn 421} are thus much more
reasonable than for \object{PKS 2155--304}.
\citet{fossati08} found the X-rays and VHE $\g$-rays to be
highly correlated, indicating that they are likely emitted from the
same part of the jet, although a highly variable external X-ray source
that is Compton scattered by a blob would also explain this.  As seen
in Fig.\ \ref{coolingMkn421}, the cooling timescale is significantly
longer than the variability timescale, as with the \object{PKS
2155--304} flare.  Thus the variability discussion in the previous
section applies to \object{Mkn 421} as well, and it is possible that
Mkn 421's variability is dominated by some other mechanism.  Analysis
of hardness-intensity diagrams could indicate the effects of
additional radiation mechanisms or variability sources
\citep[e.g.,][]{kat00,li00,boett02a}.  For example, if adiabatic
expansion dominates the energy-loss rate of electrons with
sufficiently low Lorentz factors, then the corresponding achromatic
synchrotron spectral variability should allow one to distinguish this
from frequency-dependent radiative cooling effects.

\subsection{Predictions for {\em GLAST}}

Fig.\ \ref{sensitive} shows the blazar $\nu F_\nu$ flux that would be
significantly detected with {\em GLAST} in the scanning mode as a
function of observing time (see Appendix \ref{sensitivity}).  The
predicted fluxes of \object{PKS 2155--304} shown in
Figs. \ref{SEDtv300} -- \ref{SED_dermer} have a $\nu F_\nu$ flux at 1
GeV of $\approx 8\times 10^{-10}$ ergs cm$^{-2}$ s$^{-1}$ and a $\nu
F_\nu$ spectral index  $a = 0.5$.  From Fig.\
\ref{sensitive}, we see that {\em GLAST} will significantly detect
\object{PKS 2155--304} at this flux level in less than one or two ksec
when integrating above $100$ MeV, and $\lesssim 10$ ksec when
integrating above 1 GeV. Significant detection of \object{Mkn 421}
will be achieved with {\em GLAST} in $\lesssim 30$ ksec whether
integrating above 100 MeV or 1 GeV, as can be seen from  Fig.\
\ref{Mkn421}.  For \object{PKS 2155--304}, the predicted $\nu F_\nu$
flux is quite sensitive to the assumed IBL (see Fig.\ \ref{SEDtv300}),
so if the synchrotron/SSC model is valid, then the different IBLs
could in principle be distinguished.

\subsection{Summary}

We have modeled flares in \object{PKS 2155--304} and \object{Mkn
421} with a synchrotron/SSC model.  Due to the high Doppler
factors, jet powers, and radiative cooling timescales, we find that
it is unlikely that SSC emission alone can explain the giant TeV
flares in the blazar \object{PKS 2155--304}, at least for the
one-zone approximation.  Lowering the IBL energy density lowers these
quantities considerably, but not enough to change our conclusions.
Although one-zone SSC modeling of the March 2001 \object{Mkn 421}
flare gives reasonable jet powers and Doppler factors, the long
cooling timescales cause problems for this flare as well.  Compton
scattering of photons from an external radiation source, for example,
from radiation produced in different regions of the jet, or
loosening the one-zone approximation, might remedy these problems.
The addition of external scattered radiation to this analysis
technique will be the subject of future work.


\acknowledgements 
We are grateful to L.\ Foschini for providing us with {\em Swift} data
for PKS 2155--304, and to L.\ Costamante, 
L.\ Foschini, G.\ Ghisellini, and B.\ Giebels for helpful correspondence. 
We thank the anonymous referee for comments which have improved 
this work.
The work of J.D.F. is supported by NASA {\em Swift} Guest Investigator
Grant DPR-NNG05ED411 and NASA {\em GLAST} Science Investigation
DPR-S-1563-Y, which also supported a visit by M.B. to NRL.  
C.D.D. is supported by the Office of Naval Research.

\appendix

\section{{\em GLAST} Sensitivity to Blazar Flares}
\label{sensitivity}

{\em GLAST} Large Area Telescope (LAT) sensitivity estimates to blazar
flares are updated using the latest {\em GLAST} LAT performance
parameters.\footnote{See LAT Instrument Performance at
http://www-glast.slac.stanford.edu/.}
Let $u = E$(GeV) represent photon energy in GeV. The effective area of
the {\em GLAST} LAT, accurate to better than 15\% for 70 MeV $\lesssim
E \lesssim 200$ GeV, can be written as
\begin{equation}
A(u) = A_0 \cases{\sqrt{u},& $0.07\lesssim u \lesssim 1$\cr
	1,&if $1\lesssim u \lesssim 200$,\cr}
\label{Aeff}
\end{equation}
with $A_0 = 8600$ cm$^2$.  Following the approach of 
\citet{dermer07gl}, the number of
source counts detected with the {\em GLAST} LAT above photon energy
$u$ GeV is
\begin{equation}
S(>u) = {\eta_\gamma X\Delta t f_{GeV} A_0\over E_{GeV}}\;
\left\{{2(u^{a-1/2}-1) H(1-u)\over 1-2a} + 
{[\max(1,u)]^{a-1}\over {1-a}}\right\}\;.
\label{Source}
\end{equation}
Here $\Delta t$ is the total observing time and $X $ is the
occultation factor in the scanning mode, $\eta_\gamma=0.67$ is an
acceptance cone about the source, and $E_{GeV} = 1.6\times 10^{-3}$
ergs is 1 GeV in units of ergs, provided that the blazar $\nu F_\nu$
spectrum $f_\e$ is given in units of ergs cm$^{-2}$ s$^{-1}$.  The
blazar flare spectrum is assumed to be described by a single power law
with $\nu F_\nu$ index $a$, so that $f_\e = f_{\rm GeV} u^a H(u;
u_1,u_2)$, where $f_{GeV}$ is the $\nu F_\nu$ flux at 1 GeV, and the
endpoints $u_1 < 0.07$ and $u_2 \gtrsim 200$.

The {\em GLAST} LAT point spread function is given in terms of the
angle $\theta(^\circ ) = 0.62 u^{-3/4}$ for 68\% containment, and is
accurate to better than 20\% for $0.03 < u < 40$.  The
energy-dependent solid angle corresponding to this point spread
function is therefore $\Delta \Omega(u) = \pi\theta^2 = 3.7\times
10^{-4} u^{-3/2} \equiv \omega_0 u^{-3/2}$ sr. The number of
background counts with energy $> u$ GeV is
\begin{equation}
B(>u) = A_0X\Delta t k_\gamma\omega_0 \left\{
\left( {u_1^{-\alpha_\gamma - 1}\over \alpha_\gamma}\right) H(1-u)
+
{\;2[\max(1,u)]^{-(\alpha_\gamma +0.5)}\over 1+2\alpha_\gamma}\;\right\}\;,
\label{background}
\end{equation}
using the diffuse $\gamma$-ray background measured with EGRET
\citep{sreekumar98} with photon index $\alpha_\gamma = 2.10$ and
coefficient $k_\gamma = 1.37\times 10^{-6}$ ph/(cm$^2$-s-sr).

Fig.\ \ref{sensitive} shows how bright the $\nu F_\nu$ flux has to be
for {\em GLAST} to detect at least 5 counts and at $n= 5$ $\sigma$
detection, estimated through the relation $n = S(>u)/\sqrt{2B(>u)}$
for $u = 0.1$ and $u = 1$, corresponding to $E>100$ MeV and $E > 1$
GeV, respectively. Results are shown for a flat, $a = 0$, and rising,
$a = 1/2$, $\nu F_\nu$ spectrum with $X = 0.2$.  The right-hand axis
shows the corresponding integral photon flux for the $u>0.1$, $a = 0$
case in units of $10^{-8}$ ph$(>100$ MeV)/(cm$^2$-s).  The break in
these curves represents a transition from a signal-dominated, bright
flux regime where the detection sensitivity $\propto \Delta t$ to a
background-dominated, dim flux regime where the detection sensitivity
$\propto \sqrt{\Delta t}$.


\bibliographystyle{apj}
\bibliography{references,blazar_ref}

\begin{thebibliography}{80}
\expandafter\ifx\csname natexlab\endcsname\relax\def\natexlab#1{#1}\fi

\bibitem[{{Aharonian} {et~al.}(2002)}]{aharonian02}
{Aharonian}, F. {et~al.} 2002, \aap, 393, 89

\bibitem[{{Aharonian} {et~al.}(2006)}]{aharonian06}
---. 2006, \nat, 440, 1018

\bibitem[{{Aharonian} {et~al.}(2007{\natexlab{a}})}]{aharonian07_2155}
---. 2007{\natexlab{a}}, \apjl, 664, L71

\bibitem[{{Aharonian} {et~al.}(2007{\natexlab{b}})}]{aharonian07_0229}
---. 2007{\natexlab{b}}, \aap, 475, L9

\bibitem[{{Albert} {et~al.}(2007)}]{albert07}
{Albert}, J. {et~al.} 2007, \apj, 669, 862

\bibitem[{{Atoyan} \& {Dermer}(2003)}]{atoyan03}
{Atoyan}, A.~M. \& {Dermer}, C.~D. 2003, \apj, 586, 79

\bibitem[{{Bednarek} \& {Protheroe}(1997)}]{bednarek97}
{Bednarek}, W. \& {Protheroe}, R.~J. 1997, \mnras, 292, 646

\bibitem[{{Bednarek} \& {Protheroe}(1999)}]{bednarek99}
---. 1999, \mnras, 310, 577

\bibitem[{{Begelman} {et~al.}(2008){Begelman}, {Fabian}, \&
  {Rees}}]{begelman08}
{Begelman}, M.~C., {Fabian}, A.~C., \& {Rees}, M.~J. 2008, \mnras, 384, L19

\bibitem[{{Benbow} {et~al.}(2006){Benbow}, {Costamante}, \&
  {Giebels}}]{benbow06}
{Benbow}, W., {Costamante}, L., \& {Giebels}, B. 2006, The Astronomer's
  Telegram, 867, 1

\bibitem[{{B{\l}a{\.z}ejowski} {et~al.}(2000){B{\l}a{\.z}ejowski}, {Sikora},
  {Moderski}, \& {Madejski}}]{blazejowski00}
{B{\l}a{\.z}ejowski}, M., {Sikora}, M., {Moderski}, R., \& {Madejski}, G.~M.
  2000, \apj, 545, 107

\bibitem[{{Bloom} \& {Marscher}(1996)}]{bloom96}
{Bloom}, S.~D. \& {Marscher}, A.~P. 1996, \apj, 461, 657

\bibitem[{{Blumenthal} \& {Gould}(1970)}]{blumen70}
{Blumenthal}, G.~R. \& {Gould}, R.~J. 1970, Reviews of Modern Physics, 42, 237

\bibitem[{{B{\"o}ttcher}(2007)}]{boett07}
{B{\"o}ttcher}, M. 2007, \apss, 309, 95

\bibitem[{{B{\"o}ttcher} \& {Chiang}(2002)}]{boett02a}
{B{\"o}ttcher}, M. \& {Chiang}, J. 2002, \apj, 581, 127

\bibitem[{{B\"ottcher} {et~al.}(1997){B\"ottcher}, {Mause}, \&
  {Schlickeiser}}]{boett97}
{B\"ottcher}, M., {Mause}, H., \& {Schlickeiser}, R. 1997, \aap, 324, 395

\bibitem[{{B{\"o}ttcher} {et~al.}(2002){B{\"o}ttcher}, {Mukherjee}, \&
  {Reimer}}]{boett02b}
{B{\"o}ttcher}, M., {Mukherjee}, R., \& {Reimer}, A. 2002, \apj, 581, 143

\bibitem[{{B{\"o}ttcher} {et~al.}(2003)}]{boett03}
{B{\"o}ttcher}, M. {et~al.} 2003, \apj, 596, 847

\bibitem[{{Brown} {et~al.}(1973){Brown}, {Mikaelian}, \& {Gould}}]{brown73}
{Brown}, R.~W., {Mikaelian}, K.~O., \& {Gould}, R.~J. 1973, \aplett, 14, 203

\bibitem[{{Celotti} \& {Fabian}(1993)}]{celotti93}
{Celotti}, A. \& {Fabian}, A.~C. 1993, \mnras, 264, 228

\bibitem[{{Celotti} {et~al.}(2007){Celotti}, {Ghisellini}, \&
  {Fabian}}]{celotti07}
{Celotti}, A., {Ghisellini}, G., \& {Fabian}, A.~C. 2007, \mnras, 375, 417

\bibitem[{{Costamante} {et~al.}(2007)}]{costamante07}
{Costamante}, L. {et~al.} 2007, High Energy Phenomena in Relativistic Outflows
  Workshop, Dublin, Ireland

\bibitem[{{Crusius} \& {Schlickeiser}(1986)}]{crusius86}
{Crusius}, A. \& {Schlickeiser}, R. 1986, \aap, 164, L16

\bibitem[{{Dermer}(2007{\natexlab{a}})}]{dermer07cr}
{Dermer}, C.~D. 2007{\natexlab{a}}, 30th International Cosmic Ray Conference,
  Merida, Mexico, arXiv: 0711.2804

\bibitem[{{Dermer}(2007{\natexlab{b}})}]{dermer07gl}
---. 2007{\natexlab{b}}, \apj, 659, 958

\bibitem[{{Dermer} \& {Atoyan}(2004)}]{dermer04}
{Dermer}, C.~D. \& {Atoyan}, A. 2004, \apjl, 611, L9

\bibitem[{{Dermer} \& {Schlickeiser}(1993)}]{dermer93}
{Dermer}, C.~D. \& {Schlickeiser}, R. 1993, \apj, 416, 458

\bibitem[{{Dermer} \& {Schlickeiser}(2002)}]{dermer02}
---. 2002, \apj, 575, 667

\bibitem[{{Dermer} {et~al.}(1992){Dermer}, {Schlickeiser}, \&
  {Mastichiadis}}]{dermer92}
{Dermer}, C.~D., {Schlickeiser}, R., \& {Mastichiadis}, A. 1992, \aap, 256, L27

\bibitem[{{Dondi} \& {Ghisellini}(1995)}]{dondi95}
{Dondi}, L. \& {Ghisellini}, G. 1995, \mnras, 273, 583

\bibitem[{{Falomo} {et~al.}(1993){Falomo}, {Pesce}, \& {Treves}}]{falomo93}
{Falomo}, R., {Pesce}, J.~E., \& {Treves}, A. 1993, \apjl, 411, L63

\bibitem[{{Foschini} {et~al.}(2006){Foschini}, {Tagliaferri}, {Pian},
  {Ghisellini}, {Treves}, {Maraschi}, {Tavecchio}, {di Cocco}, \&
  {Rosen}}]{foschini06b}
{Foschini}, L., {Tagliaferri}, G., {Pian}, E., {Ghisellini}, G., {Treves}, A.,
  {Maraschi}, L., {Tavecchio}, F., {di Cocco}, G., \& {Rosen}, S.~R. 2006,
  \aap, 455, 871

\bibitem[{{Foschini} {et~al.}(2007)}]{foschini07}
{Foschini}, L. {et~al.} 2007, \apjl, 657, L81

\bibitem[{{Fossati} {et~al.}(2004){Fossati}, {Buckley}, {Edelson}, {Horns}, \&
  {Jordan}}]{fossati04}
{Fossati}, G., {Buckley}, J., {Edelson}, R.~A., {Horns}, D., \& {Jordan}, M.
  2004, New Astronomy Review, 48, 419

\bibitem[{{Fossati} {et~al.}(2008)}]{fossati08}
{Fossati}, G. {et~al.} 2008, \apj, 677, 906

\bibitem[{{Gaisser}(1990)}]{gaisser90}
{Gaisser}, T.~K. 1990, {Cosmic rays and particle physics} (Cambridge and New
  York, Cambridge University Press, 1990, 292 p.)

\bibitem[{{Georganopoulos} \& {Kazanas}(2003)}]{georgan03}
{Georganopoulos}, M. \& {Kazanas}, D. 2003, \apjl, 594, L27

\bibitem[{{Ghisellini} {et~al.}(1988){Ghisellini}, {Guilbert}, \&
  {Svensson}}]{ghisellini88}
{Ghisellini}, G., {Guilbert}, P.~W., \& {Svensson}, R. 1988, \apjl, 334, L5

\bibitem[{{Ghisellini} \& {Madau}(1996)}]{ghisellini96}
{Ghisellini}, G. \& {Madau}, P. 1996, \mnras, 280, 67

\bibitem[{{Ghisellini} \& {Tavecchio}(2008)}]{ghisellini08}
{Ghisellini}, G. \& {Tavecchio}, F. 2008, MNRAS, submitted, arXiv: 0801.2569

\bibitem[{{Ghisellini} {et~al.}(2005){Ghisellini}, {Tavecchio}, \&
  {Chiaberge}}]{ghisellini05}
{Ghisellini}, G., {Tavecchio}, F., \& {Chiaberge}, M. 2005, \aap, 432, 401

\bibitem[{{Gould} \& {Schr{\'e}der}(1967)}]{gould67}
{Gould}, R.~J. \& {Schr{\'e}der}, G.~P. 1967, Physical Review, 155, 1404

\bibitem[{{Hartman} {et~al.}(1999)}]{hartman99}
{Hartman}, R.~C. {et~al.} 1999, \apjs, 123, 79

\bibitem[{{Jones}(1968)}]{jones68}
{Jones}, F.~C. 1968, Physical Review, 167, 1159

\bibitem[{{Joshi} \& {B{\"o}ttcher}(2007)}]{joshi07}
{Joshi}, M. \& {B{\"o}ttcher}, M. 2007, \apj, 662, 884

\bibitem[{{Kataoka} {et~al.}(2000){Kataoka}, {Takahashi}, {Makino}, {Inoue},
  {Madejski}, {Tashiro}, {Urry}, \& {Kubo}}]{kat00}
{Kataoka}, J., {Takahashi}, T., {Makino}, F., {Inoue}, S., {Madejski}, G.~M.,
  {Tashiro}, M., {Urry}, C.~M., \& {Kubo}, H. 2000, \apj, 528, 243

\bibitem[{{Krawczynski} {et~al.}(2002){Krawczynski}, {Coppi}, \&
  {Aharonian}}]{krawczy02}
{Krawczynski}, H., {Coppi}, P.~S., \& {Aharonian}, F. 2002, \mnras, 336, 721

\bibitem[{{Levinson}(2007)}]{levinson07}
{Levinson}, A. 2007, \apjl, 671, L29

\bibitem[{{Li} \& {Kusunose}(2000)}]{li00}
{Li}, H. \& {Kusunose}, M. 2000, \apj, 536, 729

\bibitem[{{Macomb} {et~al.}(1995)}]{macomb95}
{Macomb}, D.~J. {et~al.} 1995, \apjl, 449, L99

\bibitem[{{Mannheim} \& {Biermann}(1992)}]{mannheim92}
{Mannheim}, K. \& {Biermann}, P.~L. 1992, \aap, 253, L21

\bibitem[{{Maraschi} {et~al.}(1992){Maraschi}, {Ghisellini}, \&
  {Celotti}}]{maraschi92}
{Maraschi}, L., {Ghisellini}, G., \& {Celotti}, A. 1992, \apjl, 397, L5

\bibitem[{{M{\"u}cke} \& {Protheroe}(2001)}]{muecke01}
{M{\"u}cke}, A. \& {Protheroe}, R.~J. 2001, Astroparticle Physics, 15, 121

\bibitem[{{Mukherjee} {et~al.}(1997)}]{mukherjee97}
{Mukherjee}, R. {et~al.} 1997, \apj, 490, 116

\bibitem[{{Osterman} {et~al.}(2007){Osterman}, {Miller}, {Marshall}, {Ryle},
  {Aller}, {Aller}, \& {McFarland}}]{osterman07}
{Osterman}, M.~A., {Miller}, H.~R., {Marshall}, K., {Ryle}, W.~T., {Aller}, H.,
  {Aller}, M., \& {McFarland}, J.~P. 2007, \apj, 671, 97

\bibitem[{{Piner} \& {Edwards}(2004)}]{piner04}
{Piner}, B.~G. \& {Edwards}, P.~G. 2004, \apj, 600, 115

\bibitem[{{Piner} \& {Edwards}(2005)}]{piner05}
---. 2005, \apj, 622, 168

\bibitem[{{Press} {et~al.}(1992){Press}, {Teukolsky}, {Vetterling}, \&
  {Flannery}}]{press92}
{Press}, W.~H., {Teukolsky}, S.~A., {Vetterling}, W.~T., \& {Flannery}, B.~P.
  1992, {Numerical recipes in C. The art of scientific computing} (Cambridge:
  University Press, |c1992, 2nd ed.)

\bibitem[{{Primack} {et~al.}(2005){Primack}, {Bullock}, \&
  {Somerville}}]{primack05}
{Primack}, J.~R., {Bullock}, J.~S., \& {Somerville}, R.~S. 2005, in American
  Institute of Physics Conference Series, Vol. 745, High Energy Gamma-Ray
  Astronomy, ed. F.~A. {Aharonian}, H.~J. {V{\"o}lk}, \& D.~{Horns}, 23--33

\bibitem[{{Reimer}(2007)}]{rei07}
{Reimer}, A. 2007, \apj, 665, 1023

\bibitem[{{Sakamoto} {et~al.}(2007){Sakamoto}, {Nishijima}, {Mizukami},
  {Yamazaki}, \& {Kushida}}]{sakamoto07}
{Sakamoto}, Y., {Nishijima}, K., {Mizukami}, T., {Yamazaki}, E., \& {Kushida},
  J. 2007, \apj, submitted, ArXiv: 0712.3094

\bibitem[{{Sambruna} {et~al.}(2000)}]{sambruna00}
{Sambruna}, R.~M. {et~al.} 2000, \apj, 538, 127

\bibitem[{{Sbarufatti} {et~al.}(2006){Sbarufatti}, {Falomo}, {Treves}, \&
  {Kotilainen}}]{sbarufatti06b}
{Sbarufatti}, B., {Falomo}, R., {Treves}, A., \& {Kotilainen}, J. 2006, \aap,
  457, 35

\bibitem[{{Sikora} {et~al.}(1994){Sikora}, {Begelman}, \& {Rees}}]{sikora94}
{Sikora}, M., {Begelman}, M.~C., \& {Rees}, M.~J. 1994, \apj, 421, 153

\bibitem[{{Sreekumar} {et~al.}(1998)}]{sreekumar98}
{Sreekumar}, P. {et~al.} 1998, \apj, 494, 523

\bibitem[{{Stecker} {et~al.}(2006){Stecker}, {Malkan}, \&
  {Scully}}]{stecker06a}
{Stecker}, F.~W., {Malkan}, M.~A., \& {Scully}, S.~T. 2006, \apj, 648, 774

\bibitem[{{Stecker} {et~al.}(2007){Stecker}, {Malkan}, \& {Scully}}]{stecker07}
---. 2007, \apj, 658, 1392

\bibitem[{{Stecker} \& {Scully}(2006)}]{stecker06b}
{Stecker}, F.~W. \& {Scully}, S.~T. 2006, \apjl, 652, L9

\bibitem[{{Tagliaferri} {et~al.}(2003){Tagliaferri}, {Ravasio}, {Ghisellini},
  {Giommi}, {Massaro}, {Nesci}, {Tosti}, {Aller}, {Aller}, {Celotti},
  {Maraschi}, {Tavecchio}, \& {Wolter}}]{tagliaferri03}
{Tagliaferri}, G., {Ravasio}, M., {Ghisellini}, G., {Giommi}, P., {Massaro},
  E., {Nesci}, R., {Tosti}, G., {Aller}, M.~F., {Aller}, H.~D., {Celotti}, A.,
  {Maraschi}, L., {Tavecchio}, F., \& {Wolter}, A. 2003, \aap, 400, 477

\bibitem[{{Tavecchio} \& {Ghisellini}(2008)}]{tavecchio08}
{Tavecchio}, F. \& {Ghisellini}, G. 2008, MNRAS, in press, arXiv: 0801.0593

\bibitem[{{Tavecchio} {et~al.}(1998){Tavecchio}, {Maraschi}, \&
  {Ghisellini}}]{tavecchio98}
{Tavecchio}, F., {Maraschi}, L., \& {Ghisellini}, G. 1998, \apj, 509, 608

\bibitem[{{Tavecchio} {et~al.}(2001)}]{tavecchio01}
{Tavecchio}, F. {et~al.} 2001, \apj, 554, 725

\bibitem[{{Ulrich}(1978)}]{ulrich78}
{Ulrich}, M.-H. 1978, \apjl, 222, L3

\bibitem[{{Ulrich} {et~al.}(1975){Ulrich}, {Kinman}, {Lynds}, {Rieke}, \&
  {Ekers}}]{ulrich75}
{Ulrich}, M.-H., {Kinman}, T.~D., {Lynds}, C.~R., {Rieke}, G.~H., \& {Ekers},
  R.~D. 1975, \apj, 198, 261

\bibitem[{{Vermeulen} \& {Cohen}(1994)}]{vermeulen94}
{Vermeulen}, R.~C. \& {Cohen}, M.~H. 1994, \apj, 430, 467

\bibitem[{{Villata} {et~al.}(2004)}]{vetal04}
{Villata}, M. {et~al.} 2004, \aap, 421, 103

\bibitem[{{Weekes}(2003)}]{weekes03}
{Weekes}, T.~C. 2003, {Very high energy gamma-ray astronomy} (Very high energy
  gamma-ray astronomy, by Trevor C.~Weekes.~IoP Series in astronomy and
  astrophysics, ISBN 0750306580.~Bristol, UK: The Institute of Physics
  Publishing, 2003)

\bibitem[{{Zdziarski} \& {Lightman}(1985)}]{zdziarski85}
{Zdziarski}, A.~A. \& {Lightman}, A.~P. 1985, \apjl, 294, L79

\bibitem[{{Zhang} {et~al.}(2006){Zhang}, {Bai}, {Zhang}, {Treves}, {Maraschi},
  \& {Celotti}}]{zhang01}
{Zhang}, Y.~H., {Bai}, J.~M., {Zhang}, S.~N., {Treves}, A., {Maraschi}, L., \&
  {Celotti}, A. 2006, \apj, 651, 782

\bibitem[{{Zhang} {et~al.}(2002){Zhang}, {Treves}, {Celotti}, {Chiappetti},
  {Fossati}, {Ghisellini}, {Maraschi}, {Pian}, {Tagliaferri}, \&
  {Tavecchio}}]{zhang02}
{Zhang}, Y.~H., {Treves}, A., {Celotti}, A., {Chiappetti}, L., {Fossati}, G.,
  {Ghisellini}, G., {Maraschi}, L., {Pian}, E., {Tagliaferri}, G., \&
  {Tavecchio}, F. 2002, \apj, 572, 762

\end{thebibliography}

\clearpage
\begin{deluxetable}{crr}
\tabletypesize{\scriptsize}
\tablecaption{
Coefficients for the approximation to $R(x)$.  
}
\tablewidth{0pt}
\tablehead{
\colhead{ Coefficient } &
\colhead{ $10^{-2}<x<10^{0}$ } &
\colhead{ $10^{0}<x<10^1$ } 
}
\startdata
$A_0$  & -0.35775237  & -0.35842494 \\
$A_1$  & -0.83695385  & -0.79652041 \\
$A_2$  & -1.1449608   & -1.6113032  \\
$A_3$  & -0.68137283  & 0.26055213  \\
$A_4$  & -0.22754737  & -1.6979017  \\
$A_5$  & -0.031967334 & 0.032955035 \\
\enddata
\label{Rtable}
\end{deluxetable}
\clearpage

\clearpage
\begin{deluxetable}{ccccccccccccccc}
\rotate
\tabletypesize{\scriptsize}
\tablecaption{
Parameters giving best fit SSC spectra to PKS 2155--304 data for
different values of $t_{var}$ and IBL.  See text for details.  }
\tablewidth{0.0pt}
\tablehead{
\colhead{ Model No. } &
\colhead{ IBL } &
\colhead{ $t_{v,min}$  } &
\colhead{ $\delta_D$ } &
\colhead{ $B$ } &
\colhead{ $\zeta_B$  } &
\colhead{  $P_j$ \newline   } &
\colhead{ $\frac{P_j}{P_{j,min}}$ } &
\colhead{ $L_{tot}$  } &
\colhead{ $\frac{L_t}{P_j}$ } & 
\colhead{ $R_b^{\pr }$  } &
\colhead{ $\gp_{break}$ } &
\colhead{ $\gp_{max}$ } &
\colhead{ $K_e$ } &
\colhead{ reduced $\chi^2$ }
}
\startdata
  &  & [sec] &  & [mG] &  & [$10^{46}$ erg s$^{-1}$] &  & [$10^{42}$ erg s$^{-1}$] &  & [$10^{15}$ cm] &  &  &  &  \\
\hline
1 & S06 &   30 & $282\pm22$ & $41\pm7$ & 0.02 & 63 & 246 & 3.8 & $1.0\times10^{-4}$ & 0.23 & $1.1\times10^5$ & $1.3\times10^5$ & $9 \times 10^{38} $ & 2.8 \\
2 & S06 &  300 & $278^{+20}_{-60}$ & $5.9^{+4}_{-1}$ & 0.02 & 34 & 362 & 3.9 & $1.1\times10^{-5}$ & 2.2 & $1.1\times10^5$ & $4.8\times10^5$ & $6\times10^{40}$ & 1.8 \\
3 & S06 & 3000 & $168^{+40}_{-30}$ & $2.6^{+2}_{-1}$ & 0.03 & 14 & 167 & 1.1 & $7.6\times10^{-6}$ & 14 & $2.1\times10^5$ & $9.4\times10^5$ & $3\times10^{41}$ & 1.8 \\
\hline
4  & D07 &   30 & $230^{+40}_{-20}$ & $88\pm18$ & 0.03 & 3.8 & 145 & 6.5 & $1.7\times10^{-4}$ & 0.19 & $3.1\times10^4$ & $1.3\times10^5$ & $2\times10^{40}$ & 1.9 \\
5  & D07 &  300 & $124^{+10}_{-30}$ & $58^{+19}_{-8}$ & 0.02 & 5.0 & 92 & 2.3 &$4.6\times10^{-4}$ & 0.99 & $5.2\times10^4$ & $2.2\times10^5$ & $5\times10^{40}$ & 2.0 \\
6  & D07 & 3000 & $67\pm10$ & $35^{+15}_{-11}$ & 0.04 & 7.3 & 65 & 77 & $1.1\times10^{-3}$ & 5.4 & $9.2\times10^4$ & $4.0\times10^5$ & $9\times10^{41}$ & 2.0 \\
\hline
7  & P05 &   30 & $199^{+20}_{-30}$ & $150^{+45}_{-37}$ & 0.04 & 2.1 & 90 & 6.9& $3.3\times10^{-4}$ & 0.16 & $2.6\times10^4$ & $1.1\times10^5$ & $3\times10^{40}$ & 1.8 \\
8  & P05 &  300 & $107\pm30$ & $100\pm50$ & 0.05 & 2.8 & 58 & 24 & $8.6\times10^{-4}$ & 0.86 & $4.3\times10^4$ & $1.8\times10^5$ & $2\times10^{41}$ & 1.9 \\
9  & P05 & 3000 & $58\pm5$ & $35\pm10$ & 0.06 & 4.0 & 40 & 81 & $2.0\times10^{-3}$ & 4.7 & $7.6\times10^4$ & $3.2\times10^5$ & $9\times10^{41}$ & 1.9 \\
\enddata
\label{parameters}
\end{deluxetable}
\clearpage

\begin{deluxetable}{ccccccccccccccc}
\rotate
\tabletypesize{\scriptsize}
\tablecaption{
Parameters which give the best fit SSC spectra, for a given
$t_{v,min}$ and IBL, to Mkn 421; parameters are described in the
text.  
}
\tablewidth{0pt}
\tablehead{
\colhead{ Model No. } &  
\colhead{ IBL } &
\colhead{ $t_{v,min}$  } &
\colhead{ $\delta_D$ } &
\colhead{ $B$ } &
\colhead{ $\zeta_B$  } &
\colhead{ $P_j$ } &
\colhead{ $\frac{P_j}{P_{j,min}}$ } &
\colhead{ $L_{tot}$ } &
\colhead{ $\frac{L_t}{P_j}$ } & 
\colhead{ $R_b^{\pr}$ } &
\colhead{ $\gp_{break}$ } &
\colhead{ $\gp_{max}$ } &
\colhead{ $K_e$ } &
\colhead{ reduced $\chi^2$ }
}
\startdata
  &  & [sec] &  & [mG] &  & [$10^{45}$ erg s$^{-1}$] &  & [$10^{43}$ erg s$^{-1}$] &  & [$10^{15}$ cm] &  &  &  &  \\
\hline
10 & S06 & $10^3$ & $80\pm14$ & $48\pm9$ & 0.1 & 3.1 & 24 & 3.5 & 0.011& 3.0 & $9.3\times10^4$ & $5.1\times10^5$ & $4\times 10^{40}$ & 6.8 \\
11 & S06 & $10^4$ & $31\pm1$ & $22\pm5$ & 0.1 & 4.4 & 32 & 14 & 0.032 & 10  & $8.4\times10^5$ & $9.7\times10^5$ & $1\times10^{40}$  & 4.0 \\
\hline
12 & D07 & $10^3$ & $85\pm8$ & $35\pm22$ & 0.1 & 3.7 & 31 & 2.1 & $5.7\times10^{-3}$ & 2.5 & $1.3\times10^5$ & $5.8\times10^5$ & $2\times10^{40}$  & 4.2 \\
13 & D07 & $10^4$ & $34\pm6$ & $27\pm25$ & 0.1 & 3.5 & 26 & 10 & 0.029 & 9.9 & $7.6\times10^5$ & $8.9\times10^5$ & $1\times10^{40}$  & 3.2 \\
\hline
14 & P05 & $10^3$ & $96\pm20$ & $35\pm22$ & 0.1 & 3.7 & 24 & 1.4 & $3.8\times10^{-3}$ & 2.8 & $8.2\times10^4$ & $5.4\times10^5$ & $6\times10^{40}$  & 2.5 \\
15 & P05 & $10^4$ & $37\pm5$ & $38\pm7$ & 0.1 & 3.0 & 16 & 11 & $3.7\times10^{-3}$ & 11  & $1.9\times10^5$ & $8.5\times10^5$ & $2\times10^{41}$  & 4.4 \\
\enddata
\label{Mknparameters}
\end{deluxetable}
\clearpage

\begin{figure}
\epsscale{1.0}
\plotone{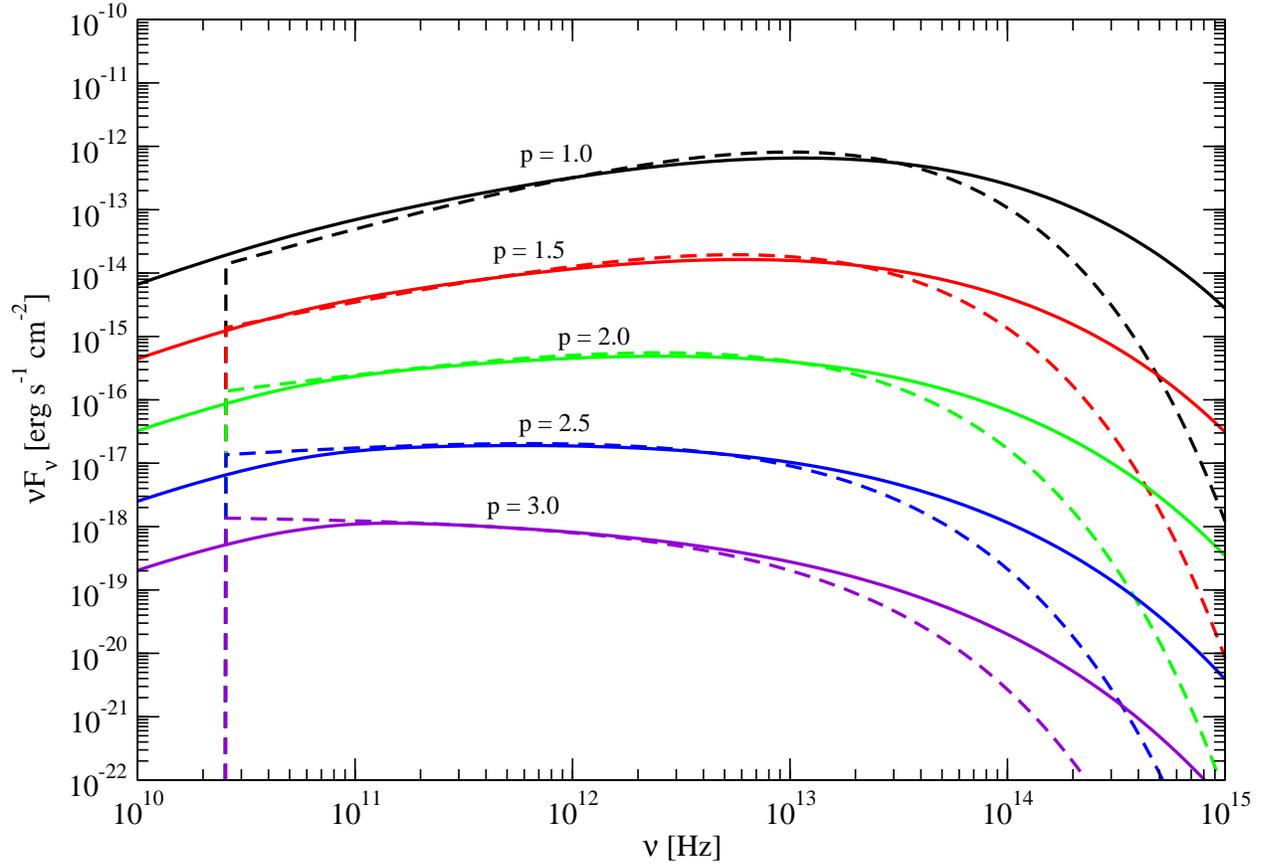}
\caption{ 
A comparison between the exact synchrotron expression (solid curves)
and the $\delta$-approximation (dashed curves) computed for various
electron spectral indices.  
}
\label{synchcompare}
\end{figure}
\clearpage

\begin{figure}
\epsscale{1.0}
\plotone{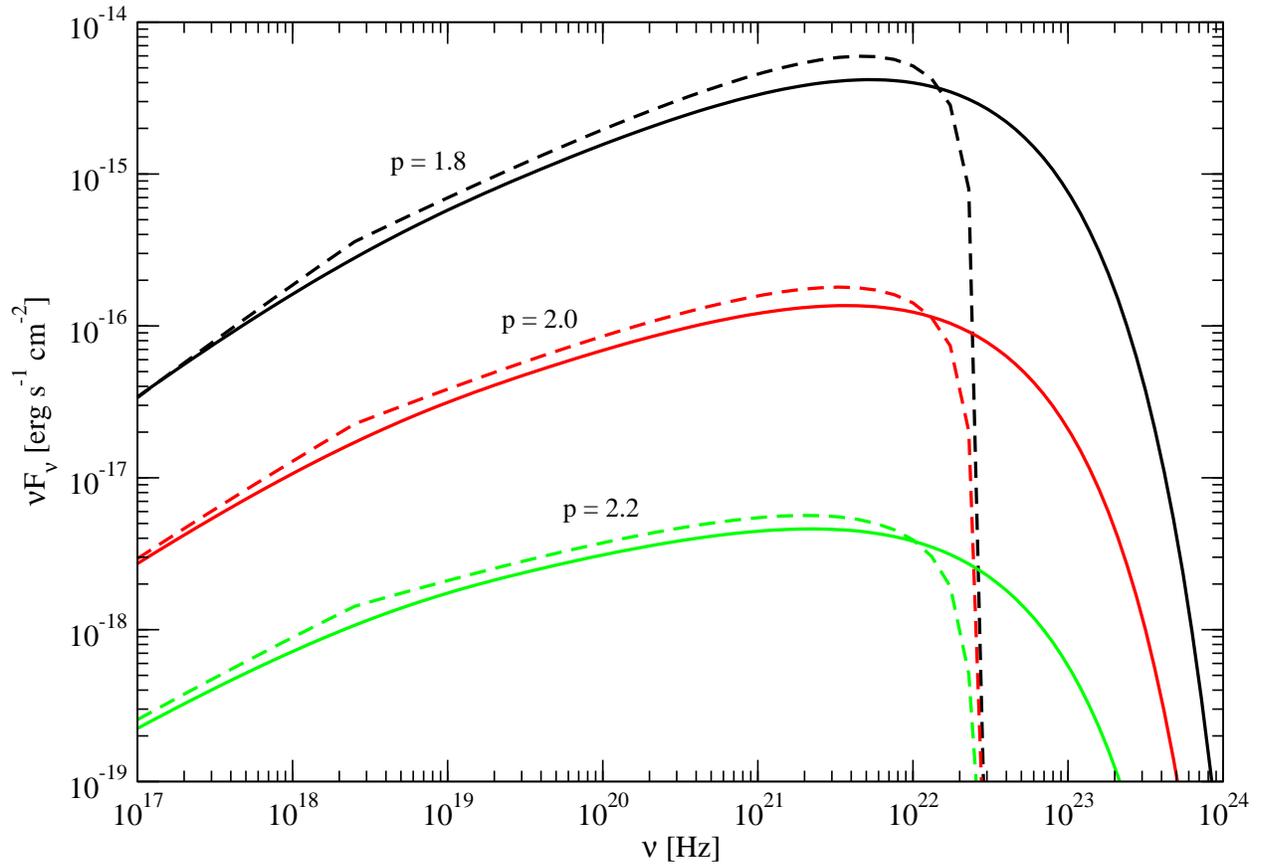}
\caption{
A comparison of SSC calculated with the full Compton cross section
(solid curves) and the Thomson cross section (dashed curves) for
various electron spectral indices.
}
\label{thomson}
\end{figure}
\clearpage

\begin{figure}
\epsscale{1.0}
\plotone{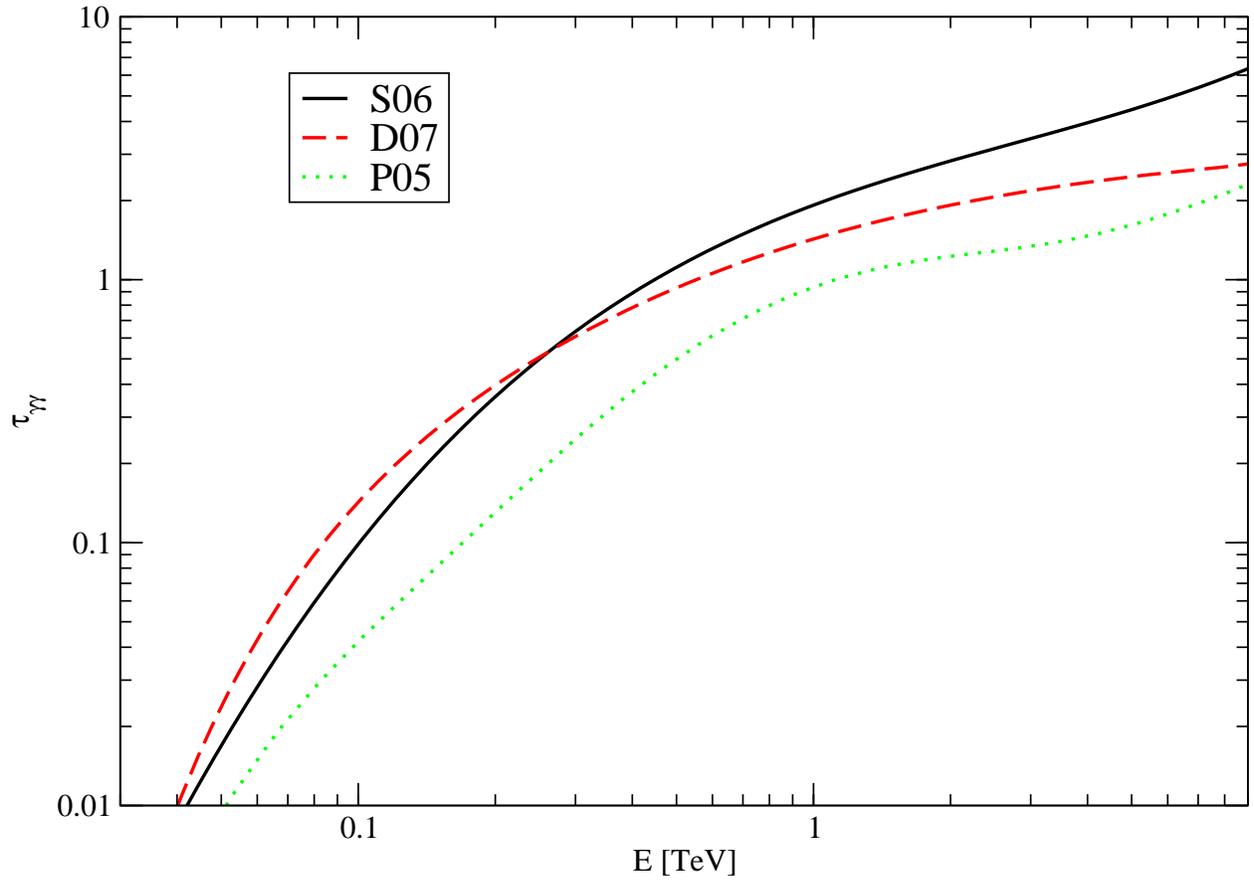}
\caption{
A comparison of the photoabsorption opacity for 
various IBL formulations for $z=0.116$, the redshift of PKS 2155--304. 
}
\label{IBLcompare}
\end{figure}
\clearpage

\begin{figure}
\epsscale{0.7}
\plotone{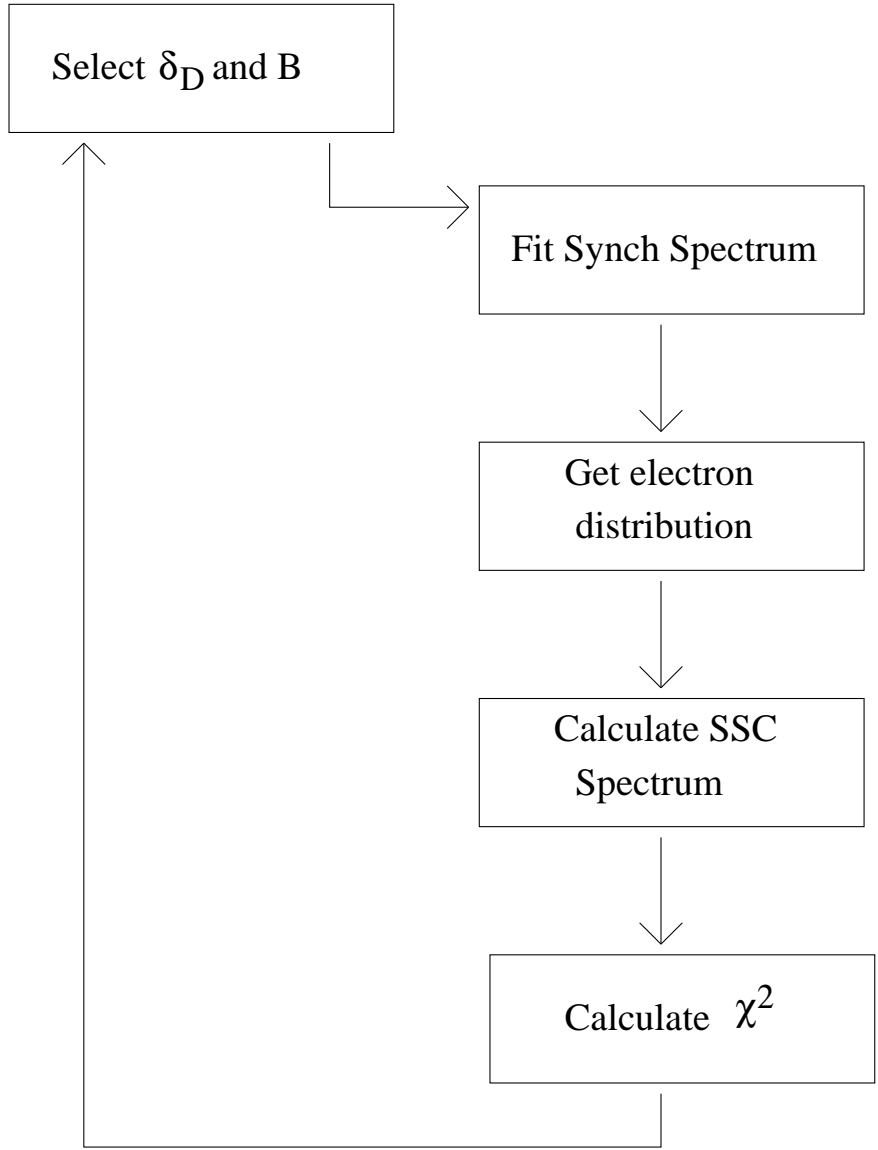}
\caption{
Diagram demonstrating the synchrotron and SSC fitting procedure.
}
\label{fitdiagram}
\end{figure}
\clearpage

\begin{figure}
\epsscale{1.0}
\plotone{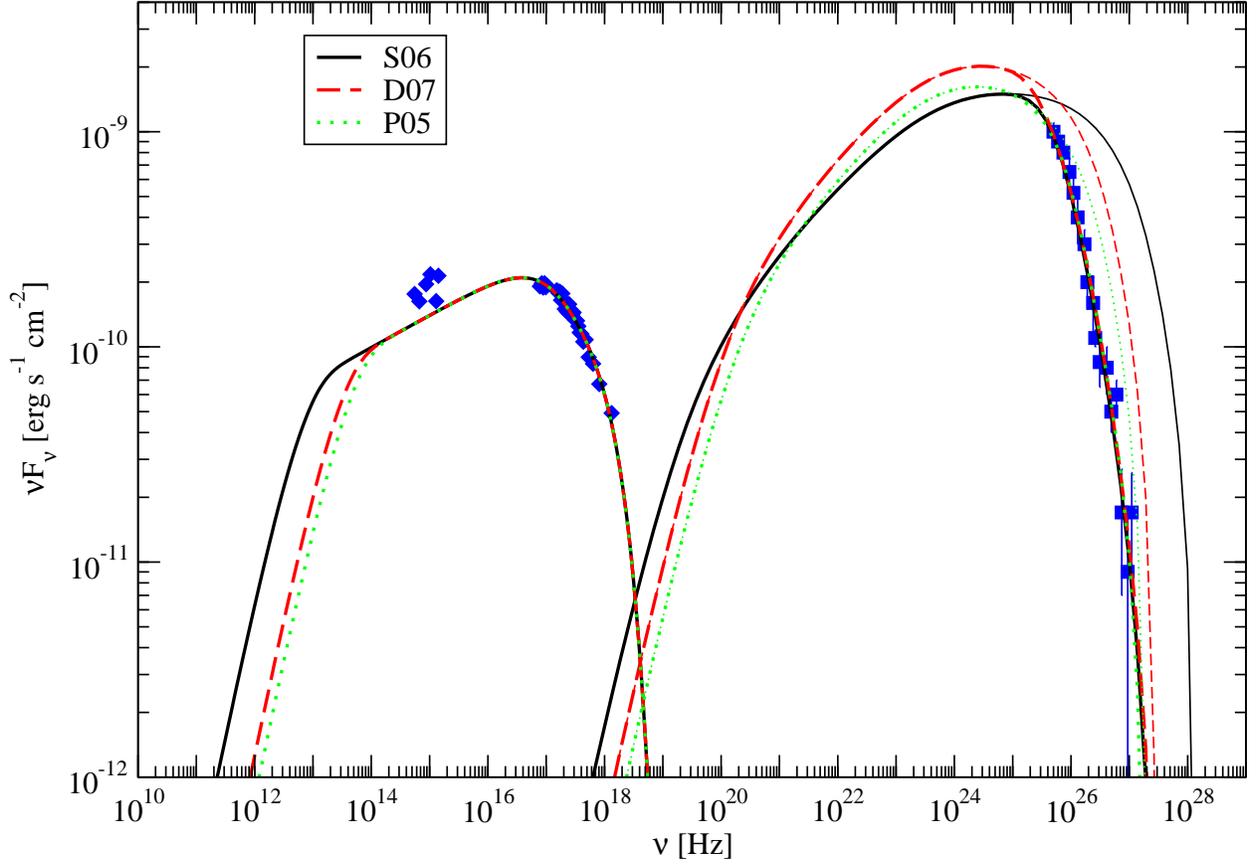}
\caption{
Various synchrotron and SSC fits for $t_{v,min}=300$ sec, for
various IBL formulations.  {\em Swift} UVOT and XRT data from PKS
2155--304 for 30 July 2006 (diamonds), as well as {\em HESS} data from
28 July 2006 (squares) are shown.  The thin curves are the unabsorbed
spectra, while the thick curves take into account $\g\g$ absorption by
jet internal radiation and the IBL.  
}
\label{SEDtv300}
\end{figure}
\clearpage

\begin{figure}
\epsscale{1.0}
\plotone{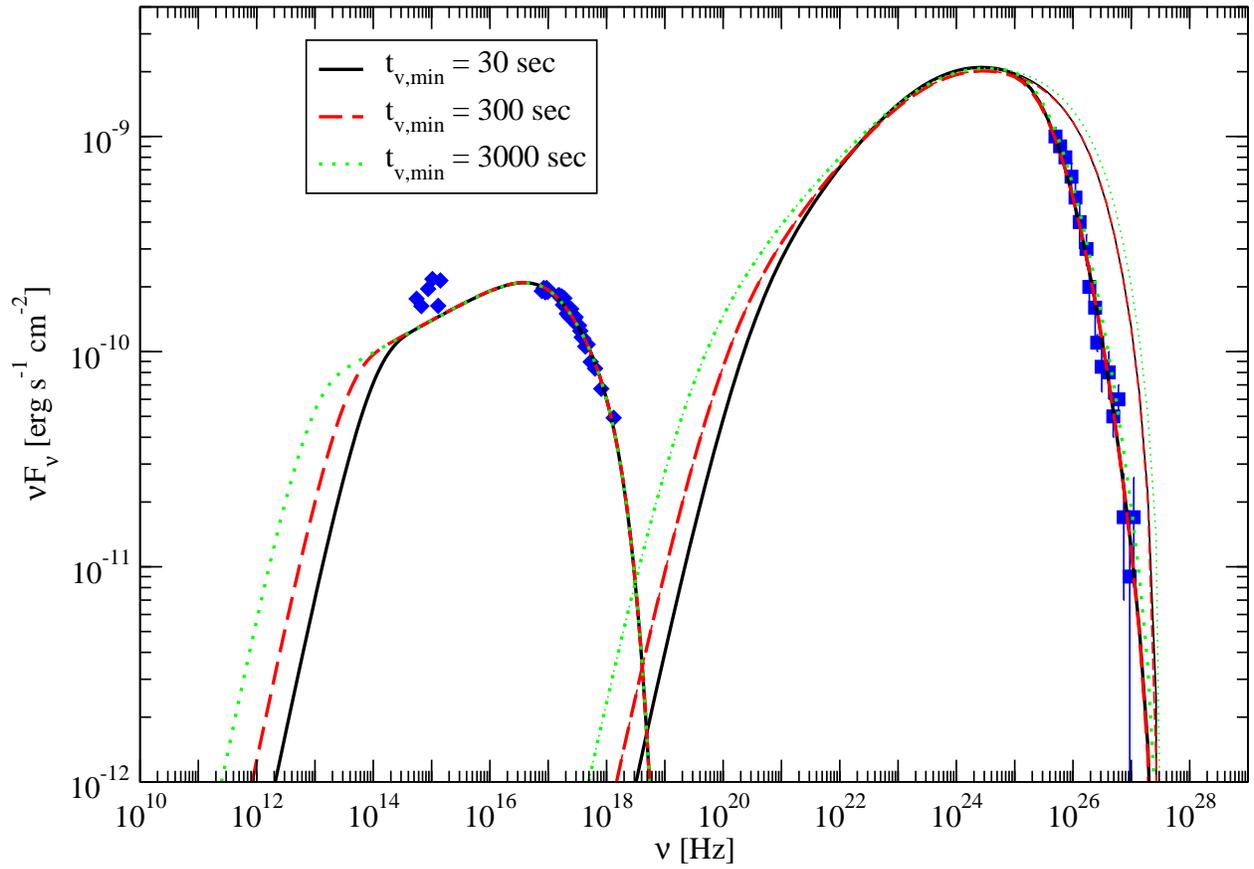}
\caption{
Similar to Fig.\ \ref{SEDtv300}
figure compares models with different $t_{v,min}$ for
the D07 IBL.  
}
\label{SED_dermer}
\end{figure}
\clearpage

\begin{figure}
\epsscale{1.0}
\plotone{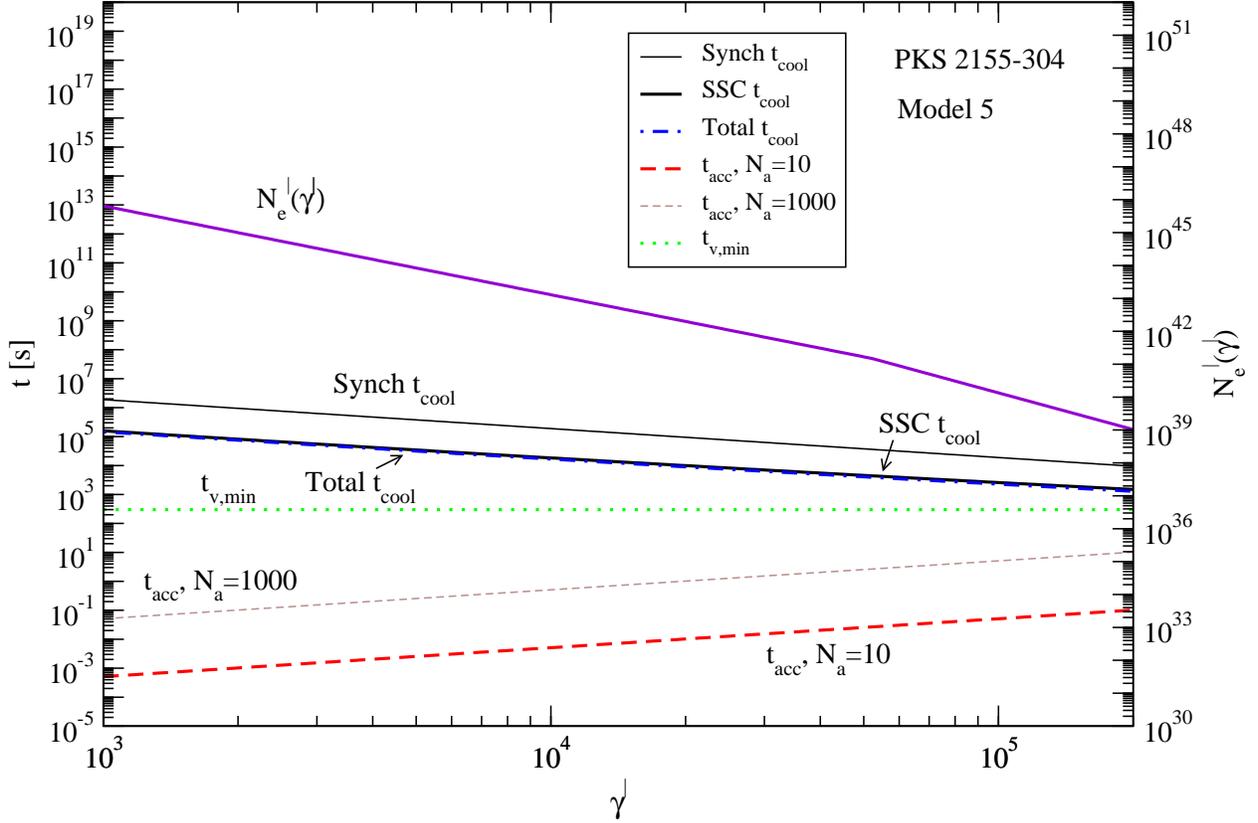}
\caption{ The cooling timescales for synchrotron (thin solid black
curve) and SSC (thick solid black curve) as a function of electron
energy for Model 5.  The total cooling time from both of these
processes is plotted as a dashed-dotted blue curve.  The acceleration
timescale is plotted as the dashed red and brown curves with $N_a = 10$
and $N_a = 1000$, respectively, and the variability timescale for this
simulation (300 s) is shown as the dotted green curve.  Also
overplotted is the form of the electron spectrum, $N_e^\prime(\gp )$
(solid violet curve).  
}
\label{cooling}
\end{figure}
\clearpage

\begin{figure}
\epsscale{1.0}
\plotone{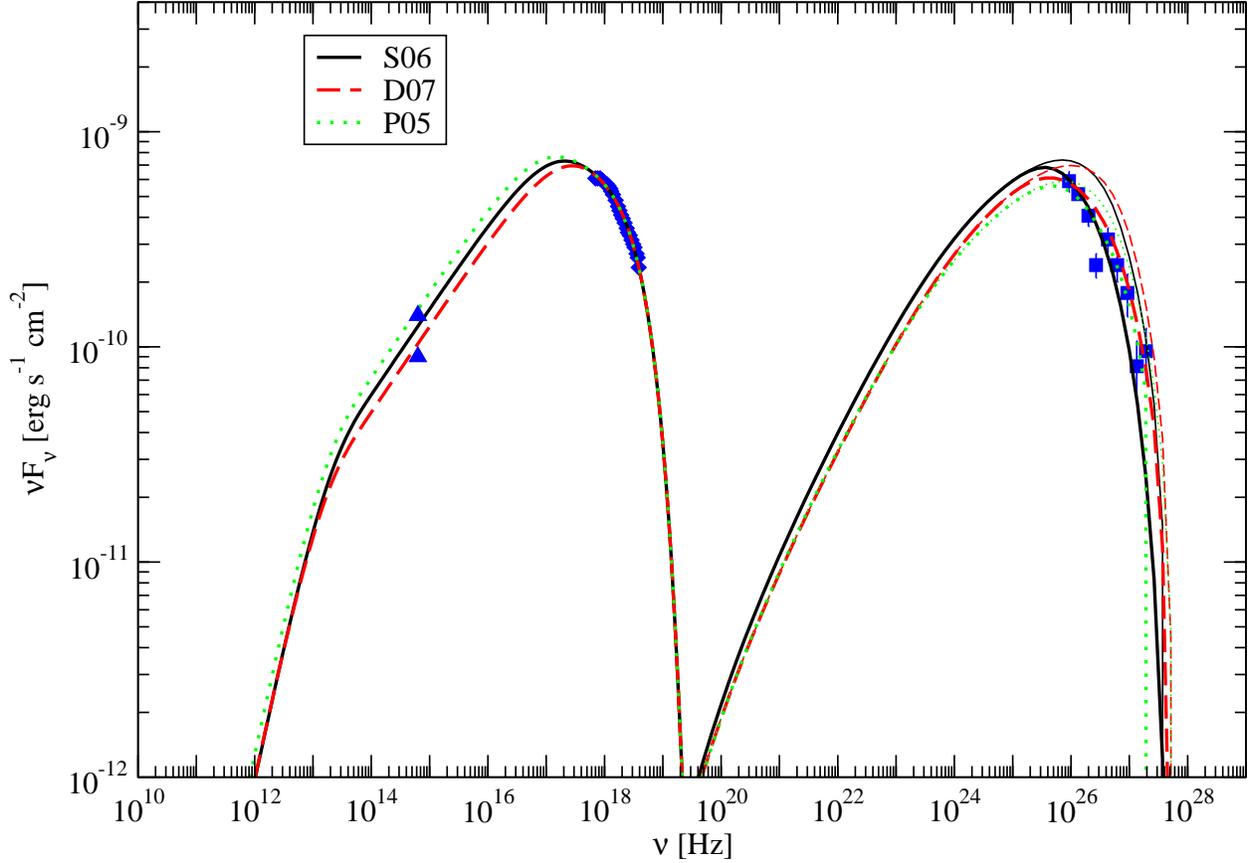}
\caption{ 
Synchrotron SSC model fits to the March 2001 flare observed by
\citet{fossati08} in Mkn 421 for $t_{v,min}=10^3$ sec.  The triangles
are limits from the Hopkins 48\arcsec\ optical observations, the
diamonds are the {\em RXTE} data, and the squares are the {\em HEGRA}
data \citep{aharonian02}.  The thin curves are the unabsorbed spectra,
while the thick curves take into account $\g\g$ absorption by jet
internal radiation and the IBL.  
}
\label{Mkn421}
\end{figure}
\clearpage

\begin{figure}
\epsscale{1.0}
\plotone{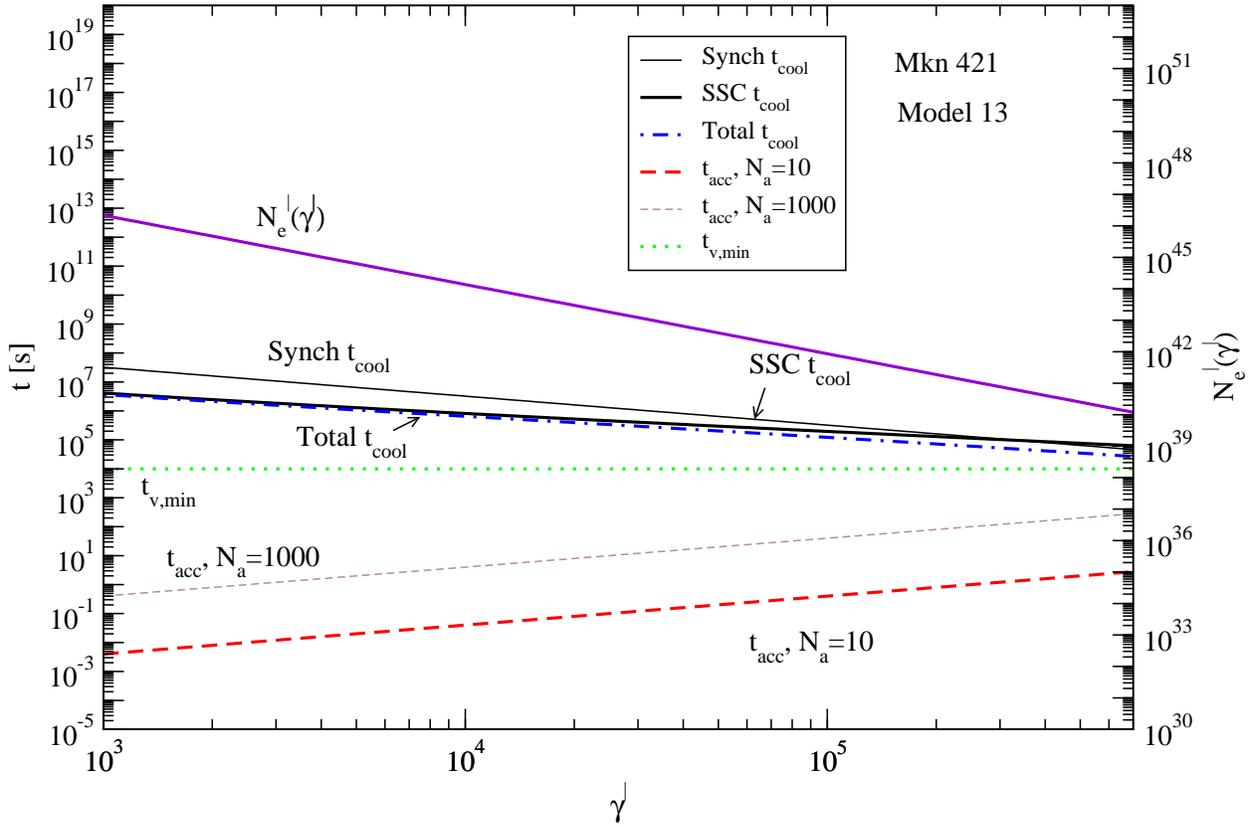}
\caption{ Similar to Fig.\ \ref{cooling} only for Model 13 
of Mkn 421. }  
\label{coolingMkn421}
\end{figure}
\clearpage

\begin{figure}
\epsscale{1.0}
\plotone{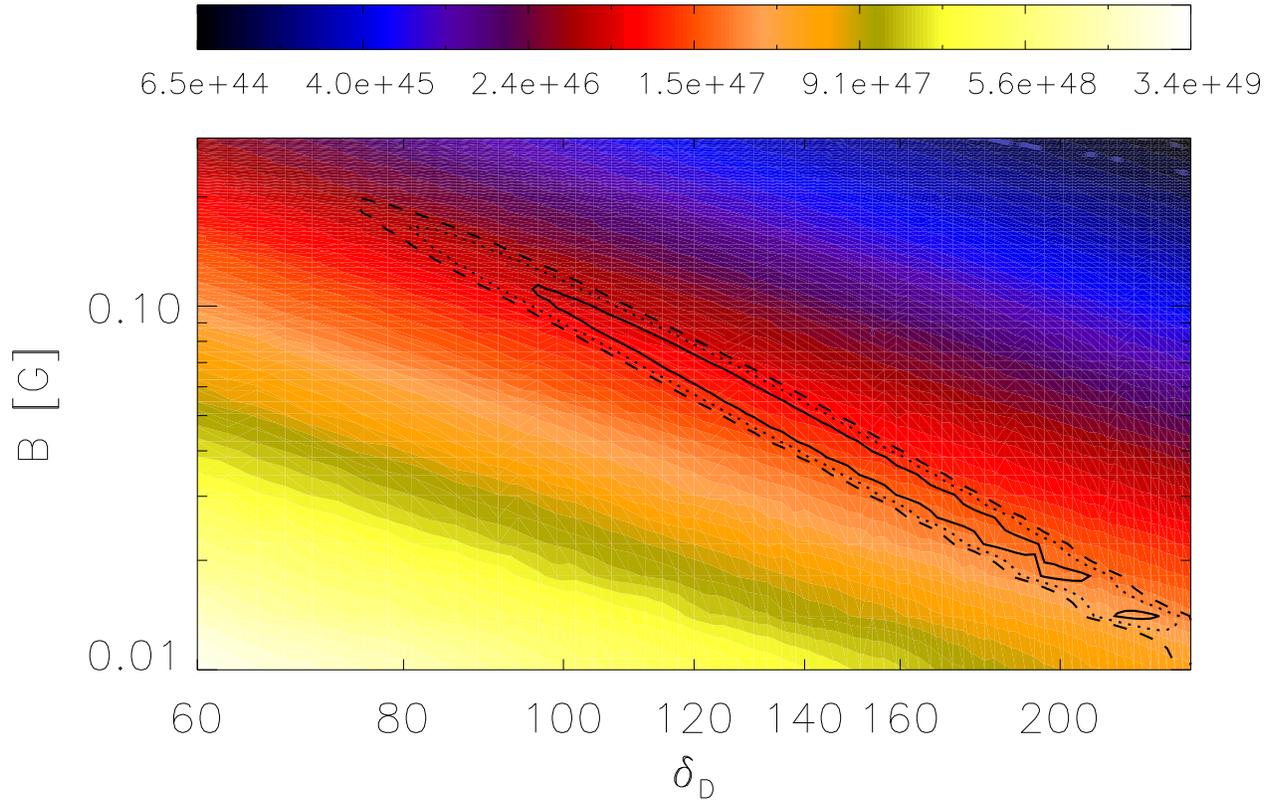}
\caption{ 
A plot of the jet power as a function of the parameters $\dD$ and $B$
for Model 5.  The color corresponding to a specific jet power is
given in the bar above, in erg s$^{-1}$.  Overplotted are the 68\%
(solid curve), 95\% (dotted curve) and 99\% (dashed curve) confidence
contours for Model 5.  
}
\label{contour}
\end{figure}
\clearpage

\begin{figure}
\epsscale{1.0}
\plotone{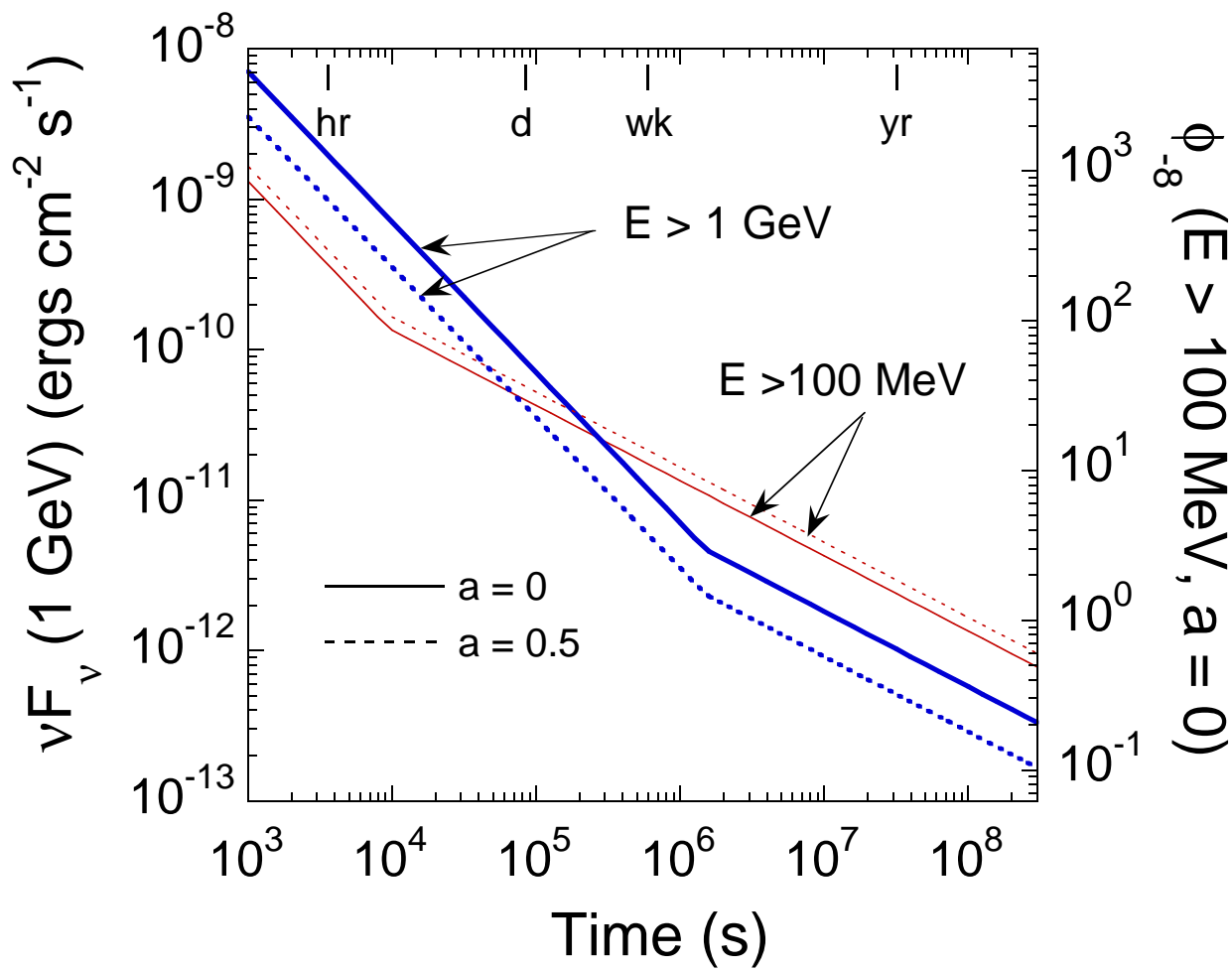}
\caption{ 
The required $\nu F_\nu$ flux at 1 GeV of a high galactic
latitude point source to be significantly detected with {\em GLAST} as
a function of total time in the scanning mode. The thick and thin
curves give the required fluxes when integrating above $100$ MeV and 1
GeV, respectively, and the solid and dotted curves show the required
fluxes for $\nu F_\nu$ spectral indices $a = 0$ and $a = 1/2$,
respectively.  
}
\label{sensitive}
\end{figure}
\clearpage

\end{document}